\documentclass[a4paper,11pt]{article}
\usepackage{jheppub} 
\usepackage{lineno}
\usepackage{mathtools}
\usepackage{braket}
\usepackage{float}
\usepackage[T1]{fontenc}

\def\bal#1\eal{\begin{align}#1\end{align}}

\title{
Generalized Glauber theorem for dark-matter axion and graviton detection
}

\author{
{\large Jakub Br\k{e}czewski$^{1}$, Ayuki Kamada$^{1}$, Akira Miyazaki$^{2}$}
\\*[20pt]
{\it \normalsize 
$^1$Institute of Theoretical Physics, Faculty of Physics, University of Warsaw, \\
ul.~Pasteura 5, PL-02-093 Warsaw, Poland \\[5pt]
$^2$Université Paris-Saclay, CNRS/IN2P3, IJCLab, Orsay, \\
15 rue G. Clémenceau FR-91405 Orsay France
} \\*[5pt]
}

\emailAdd{j.breczewski@student.uw.edu.pl, akamada@fuw.edu.pl, akira.miyazaki@ijclab.in2p3.fr}

\abstract{
We revisit the generalized Glauber theorem motivated by recent applications to dark-matter axion searches and graviton production. 
For Hamiltonian containing up to quadratic terms in annihilation and creation operators coupled to classical sources, the time-evolution operator can be factorized into displacement, squeezing, and rotation operators. 
We derive the differential equations governing time evolution of their parameters, reducing the quantum dynamics to c-number equations that can be solved analytically or numerically.
To compare another form of the time-evolution operator, Dyson series, we demonstrate the essential role of time ordering.
This formalism based on generalized Glauber theorem provides a unified description of particle production from classical backgrounds.
Axion-photon conversion in microwave haloscopes is recovered as the linear-interaction limit, while squeezed graviton production from black-hole mergers follows naturally from quadratic interactions. 
Extending the theorem to thermal initial states yields a realistic quantum description of microwave cavities used in axion experiments. 
We show that higher-order photon statistics exhibit nontrivial behavior, providing a rigorous foundation for Monte Carlo simulations of quantum-enhanced axion searches beyond heuristic noise estimates.
}

\begin{document}
\maketitle
\flushbottom

\section{Introduction}\label{sec:intro}
Particle physics and quantum optics are two landmarks of quantum theory in the 20th century.
They are both originated from the Quantum Electrodynamics and share similar formalism and terminology.
However, until recently, these two fields have been practically decoupled due to the fact that typical particle physics experiments, such as collider physics, feature particle counting and do not address quantum coherency.
For example, we do not discuss quantum coherency of two photons from a decaying Higgs boson.
In order to address quantum coherency, the detection scheme should be designed to measure the field operators and reveal quantum interferences; however, particle counting of high-energy photons is based on only local elementary processes, such as the photoelectric effect, Compton scattering, and pair creation, so that the field operators themselves are seldom considered to be experimentally measurable.
Measurement of manipulated quantum states have been a privilege of quantum optics or related fields.

The situation has been changing in the emerging research fields in the 21st century, namely dark-matter axion experiments~\cite{PhysRevLett.120.151301, PhysRevLett.124.101303, PhysRevLett.127.261803, PhysRevLett.133.051802, PhysRevX.14.031023, PhysRevD.97.092001,85c43e4df4cf4005bdbb90a5b13f9a87, PhysRevLett.132.031601, Quiskamp:2022pks, CAST:2020rlf, c749-419q} and gravitational-wave detections~\cite{PhysRevLett.116.061102, PhysRevLett.127.011103, kw5g-d732}.
In order to improve the measurement sensitivity, quantum squeezing have been implemented in gravitational-wave observatories~\cite{LIGOScientific:2011imx, LIGOScientific:2013pcc, Virgo:2019juy}. 
Parametric amplifiers in the phase-sensitive operation are developed to introduce squeezing in several experiments of axions, firstly demonstrated by the HAYSTAC collaboration~\cite{HAYSTAC:2020kwv}, in order to overcome the standard quantum limit of the quantum-limited linear amplifier chains~\cite{PhysRevD.26.1817, PhysRevD.88.035020}.
Another approach~\cite{PhysRevD.88.035020} is to introduce bolometers and single photon counting~\cite{PhysRevApplied.14.034055, Lee:2020fen, PhysRevLett.129.261801, PhysRevD.111.075022, devlin2026penningtrapsinglephotoncounter} and has recently been realized~\cite{PhysRevX.15.021031}.
Future axion searches propose the use of squeezing and/or photon counting, for example, in Ref.~\cite{PhysRevD.107.055013}.

Even the quantum statistical nature of dark-matter axion itself has recently been studied theoretically.
For example, Ref.~\cite{PhysRevD.106.043517} showed squeezing of dark-matter axion states in the galaxy by solving the Gross-Pitaevskii and Poisson equation with a quantum coherent state as an initial state.
Ref.~\cite{PhysRevD.107.063518} calculated the thermalization process of dark-matter axion by solving the quantum master equation, again with a quantum coherent state as an initial state.
Ref.~\cite{https://doi.org/10.1002/andp.202200609} considered the detection of quantum states of dark-matter axions.
It is however important to note that the standard axion cosmology argues that the dark-matter axion field is classical due to the large occupation number~\cite{sakurai}.
This may be mathematically justified by the quantum central limit theorem~\cite{PhysRevD.111.015028}.
Ref.~\cite{Ioannisian_2017} concluded that calculations based on quantum field theory~\cite{PhysRevD.45.1782} 
does not change the mean of the expected signal power while the standard deviation may be different from the classical axion detection theory.
More recent works claim that the non-classicality of axion dark matters are not detectable~\cite{bao2025intrinsicallyquantumeffectsaxion, bao2026suppressedquantumeffectsweakly}.

A crucial matter is quantum statistical nature of secondary particles generated from classical objects.
Here, one should carefully distinguish the argument on quantum statistics of dark-matter axion itself and microwave fields (i.e. secondary particles) converted via the inverse Primakoff effect.
Since the experimental apparatus measures such secondary microwaves, not directly the dark-matter axion itself,
their quantum nature of the former is more directly essential than that of the latter for experimentalists.
The same applies to the microwave fields generated from the gravitational waves (or gravitons) via inverse Gertsenshtein effect~\cite{gertsenshtein1961}.
The most fundamental aspect of such a phenomenon, namely from classical primary to quantum secondary, was firstly derived by Glauber~\cite{PhysRev.131.2766} in the field of quantum optics.
In short, a classical electric current emits quantized photons in a quantum coherent state.
Most theoretical or experimental works today are more or less {\it implicitly} assuming this Glauber theorem.

A similar but slightly more intricate problem is of recent interest in gravitons which can be potentially generated from a classical black-hole merger~\cite{Parikh:2020nrd, PhysRevD.103.044017, PhysRevLett.127.081602, kv1t-j27m}.
Because of non-linear nature of gravity, not only a linear coupling to a classical source, but also quadratic self-couplings and mode mixings inevitably exist.
Again, this problem was studied in quantum optics~\cite{Schumaker:1986tlu, Ma:1990llj}.
In general, when Hamiltonian contains up to quadratic terms of fields (or annihilation/creation operators), the time-evolution operator is given by displacement, squeeze, and rotation operators (up to an irrelevant phase factor). We dub it as generalized Glauber theorem.
On the other hand, this generalized Glauber theorem is rarely discussed {\it explicitly} in the field of particle physics.
Keeping in mind readers from the background of particle physics, we make this paper very comprehensive; we provide (almost) all derivations of useful formulas even if they can be found in the standard literature of quantum optics.
This is also why we describe them in a natural way for readers more or less familiar with Lie algebra and group (at the level of particle physics).
For example, mathematically speaking, the generalized Glauber theorem follows from the (general) Baker-Campbell-Hausdorff (BCH) theorem and that up to quadratic terms of annihilation/creation operators form a closed algebra.

To apply generalized Glauber theorem to concrete experiments of our interest in particle physics, the existence proof is not enough; one needs to find parameters of displacement, squeeze, and rotation operators.
(Ref.~\cite{Das:2025kyn} pursues a similar goal for gravitons, while this paper is aimed to be more general/comprehensive.)
Differential equation of the parameters appearing in their definitions is non-linear and complicated~\cite{Schumaker:1986tlu, Ma:1990llj}.
We introduce new parameters which follow a simple linear differential equation.
We also use the time-evolution operator to compute expectation values of low-order terms of annihilation/creation operators.
As a special case, physics of dark-matter axion is identified as a displacement operator.
We provide a general formula of expectation values of annihilation/creation operators, and also a likely useful formulation in terms of ``classical'' probability density.

The rest of this paper is organized as follows.
Section~\ref{sec:def} is to review the building blocks of quantum optics.
Some formulas may not be commonly found in the field of particle physics.
In section~\ref{sec:gla}, we discuss the generalized Glauber theorem and, in particular, derive differential equations of parameters of displacement, squeeze, and rotation operators.
The time-evolution operator of a general model Hamiltonian is given exactly without a perturbative expansion.
We discuss a formal solution called Dyson series and remark the importance of time-ordering operator.
Section~\ref{sec:app} covers an application of the generalized Glauber theorem for mixed states.
The thermal coherent states are introduced as the most realistic model of quantum states in the dark-matter axion detectors. 
The last section~\ref{sec:conclusions} is deserved for conclusions.

\section{Definitions and Properties}\label{sec:def}
\subsection{Annihilation and creation operators}
The annihilation and creation operators satisfy the following commutation relation:
\bal
[\hat{a}_i, \hat{a}_j] = [\hat{a}^\dagger_i, \hat{a}^\dagger_j] = 0 \,, \quad [\hat{a}_i, \hat{a}^\dagger_j] = \delta_{ij} \,.
\eal
Note that the annihilation and creation operators form a closed algebra with an identity operator.
Here $i$ labels different modes (position/momentum, polarization and so on), which are eigenstates of a free Hamiltonian. Hereafter we assume a finite number of (relevant) modes, say, through both infrared (box) and ultraviolet (lattice) cutoff.
This is just to guarantee that matrix operations (product, inversion and so on) are mathematically well-defined.

Since the annihilation and creation operators commute for different $i$, we can construct the whole Hilbert space by a tensor product of the sub-Hilbert space for each $i$: $\bigotimes_i |\psi_i \rangle$.
A Hermitian operator $\hat{a}^\dagger_i \hat{a}_i$ is called the number operator and satisfies the following commutation relation:
\bal
[\hat{a}^\dagger_i \hat{a}_i, \hat{a}_i] = - \hat{a}_i \,, \quad [\hat{a}^\dagger_i \hat{a}_i, \hat{a}^\dagger_i] = \hat{a}^\dagger_i \,.
\eal
It follows that the annihilation (creation) operator decreases (increases) an eigenvalue of the number operator by $-1$ ($+1$).
Since the number operator is non-negative, namely $\langle \psi_i | \hat{a}^\dagger_i \hat{a}_i | \psi_i \rangle = || \hat{a}_i | \psi_i \rangle ||^2 \geq 0$ for any $\psi$, the eigenvalue needs to be non-negative integers $n_i$:
\bal
\hat{a}^\dagger_i \hat{a}_i | n_i \rangle = n_i | n_i \rangle \,.
\eal
In particular, the zero-eigenstate of the number operator is also the zero-eigenstate of the annihilation operator:
\bal
\hat{a}_i |0_i\rangle = 0 \,.
\eal
The $n_i$-eigenstate of the number operator is given by
\bal
|n_i \rangle = \frac{1}{\sqrt{n_i!}} \hat{a}^{\dagger n_i}_i |0_i\rangle \,,
\eal
where the prefactor is chosen so that $\langle m_i | n_i \rangle = \delta_{n_i, m_i}$.

\subsection{Displacement operator}
The displacement operator is defined by the annihilation/creation operators with complex parameters $\{\alpha\}$ as
\bal
\hat{D} \left( \{\alpha\} \right) = \exp\left[ \sum_i \left( \alpha_i \hat{a}^\dagger_i - \alpha^*_i \hat{a}_i \right) \right] \,.
\eal
Since the annihilation and creation operators commute for different $i$, one can write it as a product of the displacement operator for each $i$: $\hat{D} \left( \{\alpha\} \right) = \bigotimes_i \hat{D} \left( \alpha_i \right)$ with $\hat{D} \left( \alpha_i \right) = \exp\left( \alpha_i \hat{a}^\dagger_i - \alpha^*_i \hat{a}_i  \right)$.
One can see that
\bal
\hat{D}^{-1} \left( \alpha_i \right) = \hat{D} \left( -\alpha_i \right) = \hat{D}^\dagger \left( \alpha_i \right) = \exp\left( - \alpha_i \hat{a}^\dagger_i + \alpha^*_i \hat{a}_i \right) \,,
\eal
and thus the displacement operator is unitary.
It transforms the annihilation operator as
\bal
\hat{D}^\dagger \left( \alpha_i \right) \hat{a}_i \hat{D} \left( \alpha_i \right) = \sum_{n=0}^{\infty} \frac{1}{n!} \left[ \left( - \alpha_i \hat{a}^\dagger_i + \alpha^*_i \hat{a}_i \right), \hat{a}_i \right]_{n} = \hat{a}_i + \alpha_i \,.
\eal
In the first equality, we have used the BCH lemma (see appendix~\ref{sec:identities}).
Similarly, the creation operator transforms as
\bal
\hat{D}^\dagger \left( \alpha_i \right) \hat{a}^\dagger_i \hat{D} \left( \alpha_i \right) = \hat{a}^\dagger_i + \alpha^*_i \,.
\eal
Namely, the displacement operator shifts the annihilation (creation) operator by a complex number $\alpha_i$ ($\alpha^*_i$).
One can see that this transformation preserves the commutation relations of the annihilation and creation operators.

Remember the (general) BCH theorem (see appendix~\ref{sec:identities}) and that the annihilation and creation operators form a closed algebra with an identity operator.
This leads to yet another (factorized) form of the displacement operator:
\bal
\hat{D} \left( \alpha_i \right) = \exp\left( - \frac{1}{2} \alpha^*_i \alpha_i \right) \exp\left( \alpha_i \hat{a}^\dagger_i \right) \exp \left( - \alpha^*_i \hat{a}_i \right) \,.
\eal
One can prove this by noting 
\bal
\exp\left( \alpha_i \hat{a}^\dagger_i \right) \exp \left( - \alpha^*_i \hat{a}_i \right)
= \exp\left( \alpha_i \hat{a}^\dagger_i - \alpha^*_i \hat{a}_i + \frac{1}{2} \alpha^*_i \alpha_i \right).
\eal
which follows from the (weaker) BCH theorem (see appendix~\ref{sec:identities}).
One can use it to write a coherent state in terms of number states:
\bal
| \alpha_i \rangle = \hat{D} \left( \alpha_i \right) | 0_i \rangle = \exp\left( - \frac{1}{2} \alpha^*_i \alpha_i \right) \sum_{n_i = 0} \frac{1}{\sqrt{n_i!}} \alpha_i^{n_i} | n_i \rangle \,.
\label{eq:coh_state}
\eal
The overcomplete relation follows from this:
\bal
\hat{I}_i = \sum_{n_i} | n_i \rangle \langle n_i | = \int \frac{d^2 \alpha_i}{\pi} | \alpha_i \rangle \langle \alpha_i |
\eal
where $d^2 \alpha = d {\rm Re} \alpha \, d {\rm Im} \alpha$.

\subsection{Quadratic operators}
As repeated above, the annihilation and creation operators form a closed algebra with an identity operator.
In addition, the following quadratic operators
\bal
\hat{J}_{+, ij} = - i \frac{1}{2} \hat{a}^\dagger_i \hat{a}^\dagger_j \,, \quad \hat{J}_{-, ij} = - i \frac{1}{2} \hat{a}_i \hat{a}_j \,, \quad \hat{J}_{3, ij} = \frac{1}{4} \left(\hat{a}_j \hat{a}^\dagger_i + \hat{a}^\dagger_i \hat{a}_j \right) = \frac{1}{2} \left( \hat{a}^\dagger_i \hat{a}_j + \delta_{ij} \right)  \,,
\eal
with $\hat{J}_{\pm, ij} = \hat{J}_{\pm, ji}$ (but $\hat{J}_{3, ij} \neq \hat{J}_{3, ji}$), $\hat{J}^\dagger_{\pm, ij} = - \hat{J}_{\mp, ji}$ and $\hat{J}^\dagger_{3, ij} = \hat{J}_{3, ji}$, form a closed algebra\footnote{A diagonal component ($i=j$) satisfies the complexified Lie algebra of SU(2).}:
\bal
[\hat{J}_{\pm, ij}, \hat{J}_{\pm, kl}] &= 0 \,, \notag \\
[\hat{J}_{3, ij}, \hat{J}_{3, kl}] &= \frac{1}{16} \left( - \delta_{il} \hat{a}_j \hat{a}^\dagger_k + \delta_{jk} \hat{a}_l \hat{a}_i - \delta_{il} \hat{a}_j \hat{a}^\dagger_k + \delta_{jk} \hat{a}_l \hat{a}^\dagger_i + \delta_{jk} \hat{a}^\dagger_i \hat{a}_l - \delta_{il} \hat{a}^\dagger_k \hat{a}_j + \delta_{jk} \hat{a}^\dagger_i \hat{a}_l - \delta_{il} \hat{a}^\dagger_k \hat{a}_j \right) \notag \\
&= \frac{1}{2} \left( - \delta_{il} \hat{J}_{3, kj} + \delta_{jk} \hat{J}_{3, il} \right)\,, \notag \\
[\hat{J}_{+, ij}, \hat{J}_{-, kl}] &= - \frac{1}{4} \left(- \delta_{jk} \hat{a}^\dagger_i \hat{a}_l - \delta_{ik} \hat{a}^\dagger_j \hat{a}_l - \delta_{jl} \hat{a}_k \hat{a}^\dagger_i - \delta_{il} \hat{a}_k \hat{a}^\dagger_j \right) \notag \\
&= \frac{1}{2} \left( \delta_{ik} \hat{J}_{3, jl} + \delta_{jk} \hat{J}_{3, il} + \delta_{il} \hat{J}_{3, jk} + \delta_{jl} \hat{J}_{3, ik} \right)\,, \notag \\
[\hat{J}_{3, ij}, \hat{J}_{+, kl}] & = - i \frac{1}{8} \left( \delta_{jk} \hat{a}^\dagger_i \hat{a}^\dagger_l + \delta_{jl} \hat{a}^\dagger_k \hat{a}^\dagger_i + \delta_{jk} \hat{a}^\dagger_i \hat{a}^\dagger_l + \delta_{jl} \hat{a}^\dagger_k \hat{a}^\dagger_i \right) \notag \\
&= \frac{1}{2} \delta_{jk} \hat{J}_{\pm, il} + \frac{1}{2} \delta_{jl} \hat{J}_{\pm, ik}  \,.
\eal
Here we have used the identity $[\hat{X} \hat{Y}, \hat{Z} \hat{W}] = \hat{X} [\hat{Y}, \hat{Z}] \hat{W} + [\hat{X}, \hat{Z}] \hat{Y} \hat{W} + \hat{Z} \hat{X} [\hat{Y}, \hat{W}] + \hat{Z} [\hat{X}, \hat{W}] \hat{Y}$ and symmetric property of $\hat{J}_{\pm, ij}$.

The above two algebras are actually a sub-algebra of a more general algebra that consists of up to quadratic operators with
\bal
&[\hat{J}_{+, ij}, \hat{a}^\dagger_{k}] = [\hat{J}_{-, ij}, \hat{a}_{k}] = 0 \,,
\notag \\
&[\hat{J}_{+, ij}, \hat{a}_{k}] = \frac{i}{2} \left( \delta_{ik} \hat{a}^\dagger_{j} + \delta_{jk} \hat{a}^\dagger_{i} \right) \,, \quad [\hat{J}_{-, ij}, \hat{a}^\dagger_{k}] = - \frac{i}{2} \left( \delta_{ik} \hat{a}_{j} + \delta_{jk} \hat{a}_{i} \right) \,, \notag \\
&[\hat{J}_{3, ij}, \hat{a}^\dagger_{k}] = \frac{1}{2} \delta_{jk} \hat{a}^\dagger_{i} \,, \quad [\hat{J}_{3, ij}, \hat{a}_{k}] = - \frac{1}{2} \delta_{ik} \hat{a}_{j} \,.
\eal
It leads to generalized Glauber theorem as discussed below.

\subsection{Rotation operator}
The rotation operator is defined by the quadratic operators with complex parameters $\{\phi\}$ as
\bal
\hat{P} \left( \{\phi\} \right) = \exp\left( - i \sum_{i,j} \phi_{ij} \hat{a}^\dagger_i \hat{a}_j \right) \,,
\eal
where $\phi_{ij} = \phi^*_{ji}$ or $\phi = \phi^\dagger$ in the matrix notation.
Since the rotation operator in general mixes different modes, it is not possible to rewrite it as a product of the squeeze operator for each $i$.
An exception is when $\phi$ is diagonal: $\phi_{ij} = \delta_{ij} \phi_i$. Then $\hat{P} \left( \{\phi\} \right) = \bigotimes_i \hat{P} \left( \phi_i \right)$ with $\hat{P} \left( \phi_i \right) = \exp\left( - i \phi_i \hat{a}^\dagger_i \hat{a}_i \right)$.
One can see that
\bal
\hat{P}^{-1} \left( \{\phi\} \right) = \hat{P} \left( \{-\phi\} \right) = \hat{P}^\dagger \left( \{\phi\} \right) = \exp\left( i \sum_{i,j} \phi_{ij} \hat{a}^\dagger_i \hat{a}_j  \right) \,,
\eal
and thus the rotation operator is unitary.
It transforms the annihilation operator as
\bal
\hat{P}^\dagger \left( \{\phi\} \right) \hat{a}_i \hat{P} \left( \{\phi\} \right) = \sum_{n=0}^{\infty} \frac{1}{n!} \left[i \sum_{j,k} \phi_{jk} \hat{a}^\dagger_j \hat{a}_k, \hat{a}_i \right]_{n} = \sum_j \sum_{n=0}^{\infty} \frac{1}{n!} \left[(- i \phi)^n\right]_{ij} \hat{a}_j = \sum_j\left( e^{- i \phi} \right)_{ij} \hat{a}_j \,.
\eal
Similarly, the creation operator transforms as
\bal
\hat{P}^\dagger \left( \{\phi\} \right) \hat{a}^\dagger_i \hat{P} \left( \{\phi\} \right) = \sum_j \left( e^{i \phi^*} \right)_{ij} \hat{a}^\dagger_j \,.
\eal
Namely, the rotation operator leads to a unitary transformation of (indices of) the annihilation (creation) operators.
By using this unitarity, one can see that this transformation preserves the commutation relations of the annihilation and creation operators.

\subsection{Squeeze operator}
The squeeze operator is defined by the quadratic operators with complex parameters $\{\zeta\}$ as
\bal
\hat{S} \left( \{\zeta\} \right) = \exp\left[ \frac{1}{2} \sum_{i,j} \left( \zeta^*_{ij} \hat{a}_i \hat{a}_j -  \zeta_{ij} \hat{a}^\dagger_i \hat{a}^\dagger_j \right) \right] \,,
\eal
where $\zeta_{ij} = \zeta_{ji}$ or $\zeta = \zeta^T$ in the matrix notation.
Since the squeeze operator in general mixes different modes, it is not possible to rewrite it as a product of the squeeze operator for each $i$.
An exception is when $\zeta$ is diagonal: $\zeta_{ij} = \delta_{ij} \zeta_i$. Then $\hat{S} \left( \{\zeta\} \right) = \bigotimes_i \hat{S} \left( \zeta_i \right)$ with $\hat{S} \left( \zeta_i \right) = \exp\left[ \frac{1}{2} \left( \zeta^*_i \hat{a}_i \hat{a}_i -  \zeta_{i} \hat{a}^\dagger_i \hat{a}^\dagger_i \right) \right]$
One can see that
\bal
\hat{S}^{-1} \left( \{\zeta\} \right) = \hat{S} \left( \{-\zeta\} \right) = \hat{S}^\dagger \left( \{\zeta\} \right) = \exp\left( - \frac{1}{2} \sum_{i,j} \left( \zeta^*_{ij} \hat{a}_i \hat{a}_j -  \zeta_{ij} \hat{a}^\dagger_i \hat{a}^\dagger_j \right) \right) \,,
\eal
and thus the squeeze operator is unitary.
It transforms the annihilation operator as
\bal
\hat{S}^\dagger \left( \{\zeta\} \right) \hat{a}_i \hat{S} \left( \{\zeta\} \right) & = \sum_{n=0}^{\infty} \frac{1}{n!} \left[ - \frac{1}{2} \sum_{j, k} \left( \zeta^*_{jk} \hat{a}_j \hat{a}_k -  \zeta_{jk} \hat{a}^\dagger_j \hat{a}^\dagger_k \right), \hat{a}_i \right]_{n} \notag \\
& = \sum_{j} \left( \sum_{n={\rm even}} \frac{1}{n!}  \left[(\zeta \zeta^*)^{n/2}\right]_{i j} \hat{a}_j  - \sum_{n={\rm odd}} \frac{1}{n!} \left[(\zeta \zeta^*)^{(n-1)/2} \zeta \right]_{ij} \hat{a}^\dagger_j \right) \notag \\ 
&= \sum_{j} \left( \mu^*_{ij} \hat{a}_j - \nu_{ij} \hat{a}^\dagger_j \right)\,.
\label{eq:squeezed_a}
\eal
In the first equality, we have used the BCH lemma (see appendix~\ref{sec:identities}), while in the second equality, we used
\bal
\left[ - \frac{1}{2} \sum_{j,k} \left( \zeta^*_{jk} \hat{a}_j \hat{a}_k - \zeta_{jk} \hat{a}^\dagger_j \hat{a}^\dagger_k \right), \hat{a}_i \right]_{n} 
&= \sum_j \left[ (\zeta \zeta^*)^{n/2} \right]_{i j} \hat{a}_j  \quad \text{for $n=$ even,} \notag \\
&= - \sum_j \left[ (\zeta \zeta^*)^{(n-1)/2} \zeta \right]_{i j} \hat{a}^\dagger_j \quad \text{for $n=$ odd,}
\eal
which one can prove by mathematical induction.
Here in the matrix notation
\bal
\mu = \cosh \left( [\zeta^* \zeta]^{1/2} \right) \,, \quad \nu = \sinh \left( [\zeta \zeta^*]^{1/2} \right) (\zeta \zeta^*)^{-1/2} \zeta = \zeta (\zeta^* \zeta)^{-1/2} \sinh \left( [\zeta^* \zeta]^{1/2} \right) \,,
\eal
satisfy
\bal
\mu = \mu^\dagger \,, \quad \nu = \nu^T \,, \quad \mu^2 - \nu^* \nu = I \,, \quad \mu^* \nu - \nu \mu = 0 \,.
\label{eq:muandnu}
\eal
Similarly, the creation operator transforms as
\bal
\hat{S}^\dagger \left( \{\zeta\} \right) \hat{a}^\dagger_i \hat{S} \left( \{\zeta\} \right) = \sum_{j} \left( \mu_{ij} \hat{a}^\dagger_j - \nu^*_{ij} \hat{a}_j \right) \,.
\label{eq:squeezed_adagger}
\eal
Namely, the squeeze operator takes a linear combination of the annihilation and creation operators.
By noting \eqref{eq:muandnu}, one can see that this transformation preserves the commutation relations of the annihilation and creation operators.

Remember the (general) BCH theorem and that the quadratic operators form a closed algebra. 
This leads to yet another (factorized) form of the squeeze operator (throughout this manuscript, a primed operator denotes another form of an unprimed operator, which is shown to be equivalent to the original):
\bal
\hat{S}'(\{\zeta\}) &= \exp\left( -\frac{1}{2} \sum_{i,j} \tilde{\zeta}_{ij} \hat{a}^\dagger_i \hat{a}^\dagger_j \right) \exp\left[ - \frac{1}{2}\sum_{i,j} \tilde{\phi}_{ij} \left(\hat{a}^\dagger_i \hat{a}_j + \hat{a}_j \hat{a}^\dagger_i \right) \right] \exp\left( \frac{1}{2} \sum_{i,j} \tilde{\zeta}^*_{ij} \hat{a}_i \hat{a}_j \right) \notag \\
&= \exp\left( \frac{1}{2} \sum_{i,j} \tilde{\zeta}^*_{ij} \hat{a}_i \hat{a}_j \right) \exp\left[ \frac{1}{2} \sum_{i,j} \tilde{\phi}_{ij} \left(\hat{a}^\dagger_i \hat{a}_j + \hat{a}_j \hat{a}^\dagger_i \right) \right] \exp\left( - \frac{1}{2} \sum_{i,j} \tilde{\zeta}_{ij} \hat{a}^\dagger_i \hat{a}^\dagger_j \right) \,.
\eal
$\hat{S}'^{-1}(\{\zeta\}) =  \hat{S}'^\dagger(\{\zeta\})$ follows from the second line.
One can find new parameters $\{\tilde{\zeta}\}$ and $\{\tilde{\phi}\}$ in the following way.
By noting
\bal
& \exp\left( \frac{1}{2} \sum_{j,k} \tilde{\zeta}_{jk} \hat{a}^\dagger_j \hat{a}^\dagger_k \right) \hat{a}_i \exp\left( -\frac{1}{2} \sum_{j,k} \tilde{\zeta}_{jk} \hat{a}^\dagger_j \hat{a}^\dagger_k \right) = \hat{a}_i - \sum_j \tilde{\zeta}_{ij} \hat{a}'^\dagger_j \,, \notag \\
& \exp\left( \frac{1}{2} \sum_{j,k} \tilde{\zeta}_{jk} \hat{a}^\dagger_j \hat{a}^\dagger_k \right) \hat{a}^\dagger_i \exp\left( -\frac{1}{2} \sum_{j,k} \tilde{\zeta}_{jk} \hat{a}^\dagger_j \hat{a}^\dagger_k \right) = \hat{a}^\dagger_i \,, \notag \\
& \exp\left[ \frac{1}{2} \sum_{j, k} \tilde{\phi}_{jk} \left(\hat{a}^\dagger_j \hat{a}_k + \hat{a}_k \hat{a}^\dagger_j \right) \right] \hat{a}_i \exp\left[- \frac{1}{2} \sum_{j, k} \tilde{\phi}_{jk} \left(\hat{a}^\dagger_j \hat{a}_k + \hat{a}_k \hat{a}^\dagger_j \right) \right] = \sum_j \left( e^{-\tilde{\phi}^T} \right)_{ij} \hat{a}_j \,, \notag \\
& \exp\left[ \frac{1}{2} \sum_{j, k} \tilde{\phi}_{jk} \left(\hat{a}^\dagger_j \hat{a}_k + \hat{a}_k \hat{a}^\dagger_j \right) \right] \hat{a}^\dagger_i \exp\left[- \frac{1}{2} \sum_{j, k} \tilde{\phi}_{jk} \left(\hat{a}^\dagger_j \hat{a}_k + \hat{a}_k \hat{a}^\dagger_j \right) \right] = \sum_j \left( e^{\tilde{\phi}} \right)_{ij} \hat{a}^\dagger_j \,, \notag \\
& \exp\left( - \frac{1}{2} \sum_{j,k} \tilde{\zeta}^*_{jk} \hat{a}_j \hat{a}_k \right) \hat{a}_i \exp\left( \frac{1}{2} \sum_{j,k} \tilde{\zeta}^*_{jk} \hat{a}_j \hat{a}_k \right) = \hat{a}_i \,, \notag \\
& \exp\left( - \frac{1}{2} \sum_{j,k} \tilde{\zeta}^*_{jk} \hat{a}_j \hat{a}_k \right) \hat{a}^\dagger_i \exp\left( \frac{1}{2} \sum_{j,k} \tilde{\zeta}^*_{jk} \hat{a}_j \hat{a}_k \right) = \hat{a}^\dagger_i - \sum_j \tilde{\zeta}^*_{ij} \hat{a}_j \,,
\eal
one finds
\bal
& \hat{S}'^\dagger(\{\zeta\}) \hat{a}_i \hat{S}'(\{\zeta\}) = \sum_j \left[ \left( e^{-\tilde{\phi}^T} + \zeta' e^{\tilde{\phi}} \tilde{\zeta}^* \right)_{ij} \hat{a}_j - \left( \tilde{\zeta} e^{\tilde{\phi}} \right)_{ij} \hat{a}^\dagger_j \right] = \sum_j \left[ \left( e^{\tilde{\phi}^T}\right)_{ij} \hat{a}_j - \left( e^{\tilde{\phi}^T} \tilde{\zeta} \right)_{ij} \hat{a}^\dagger_j \right] \,,  \notag \\
& \hat{S}'^\dagger(\{\zeta\}) \hat{a}^\dagger_i \hat{S}'(\{\zeta\}) = \sum_j \left[ \left( e^{\tilde{\phi}}\right)_{ij} \hat{a}^\dagger_j - \left( e^{\tilde{\phi}} \tilde{\zeta}^* \right)_{ij} \hat{a}_j \right] = \sum_j \left[ \left( e^{-\tilde{\phi}} + \tilde{\zeta}^* e^{\tilde{\phi}^T} \tilde{\zeta} \right)_{ij} \hat{a}^\dagger_j - \left( \tilde{\zeta}^* e^{\tilde{\phi}^T} \right)_{ij} \hat{a}_j \right] \,.
\eal
For these transformations to be the same as for \eqref{eq:squeezed_a} and \eqref{eq:squeezed_adagger}, the followings should hold:
\bal
\mu = e^{\tilde{\phi}} \,, \quad \tilde{\zeta} = (\mu^*)^{-1} \nu \,,
\eal
in the matrix notation. By using \eqref{eq:muandnu}, one can see that these relations are consistent with $\tilde{\phi} = \tilde{\phi}^\dagger$ and $\tilde{\zeta} = \tilde{\zeta}^T$, and $\hat{S}'^{-1}(\{\zeta\}) = \hat{S}'(\{-\zeta\}) =  \hat{S}'^\dagger(\{\zeta\})$.

In summary, since they transform annihilation and creation operators in the same way, $\hat{S}'(\{\zeta\}) = \hat{S}(\{\zeta\})$ up to a phase.
Furthermore, since the algebra of the quadratic operators does not contain a center (unlike the algebra of the annihilation/creation operators), there should not be a phase.
It is easy to rewrite $\hat{S}'(\{\zeta\})$ also as
\bal
\hat{S}'(\{\zeta\}) &= e^{- \frac{1}{2} {\rm tr} \tilde{\phi}} \exp\left( -\frac{1}{2} \sum_{i,j} \tilde{\zeta}_{ij} \hat{a}^\dagger_i \hat{a}^\dagger_j \right) \exp\left[ - \sum_{i,j} \tilde{\phi}_{ij} \hat{a}^\dagger_i \hat{a}_j \right] \exp\left( \frac{1}{2} \sum_{i,j} \tilde{\zeta}^*_{ij} \hat{a}_i \hat{a}_j \right) \notag \\
&= e^{\frac{1}{2} {\rm tr} \tilde{\phi}} \exp\left( \frac{1}{2} \sum_{i,j} \tilde{\zeta}^*_{ij} \hat{a}_i \hat{a}_j \right) \exp\left[ \sum_{i,j} \tilde{\phi}_{ij} \hat{a}^\dagger_i \hat{a}_j  \right] \exp\left( - \frac{1}{2} \sum_{i,j} \tilde{\zeta}_{ij} \hat{a}^\dagger_i \hat{a}^\dagger_j \right) \,,
\eal
where $e^{{\rm tr} \tilde{\phi}} = \det \left( e^{\tilde{\phi}} \right) = \det(\mu)$.
One can use this form to write a squeezed state ${\hat S}' (\{\zeta\}) |0 \rangle$ in terms of number states~\cite{Qin:2001mms}, though we do not go into it here.
The above derivation of this form is more important for our purpose, since we follow the same strategy in the generalized Glauber theorem: namely, (general) BCH theorem and closed algebra to determine a possible form of the operator; and then transformation of the annihilation/creation operators to find the parameters appearing in the operator.

\section{Generalized Glauber theorem}\label{sec:gla}
It is shown by Glauber~\cite{PhysRev.131.2766} that time evolution of a harmonic oscillator in the presence of interaction involving a linear combination of the annihilation (creation) operator is given by the displacement operator up to a phase.
Mathematically speaking, this is because of the (general) BCH theorem and that the annihilation and creation operators form a closed algebra with an identity operator.
It physically means that photons, produced via a linear coupling to a classical source (current), follow a coherent state.

This Glauber theorem is extended to interaction involving terms up to quadratic operators.
The time-evolution operator is given by a product of the displacement, squeeze and rotation operators up to a phase factor (generalized Glauber theorem)~\cite{Schumaker:1986tlu, Ma:1990llj}.
Mathematically speaking, this is because up to the quadratic operators form a closed algebra.
Physically speaking, it means that photons, produced via linear and quadratic couplings to classical sources, follow a displaced squeezed state.

On the other hand, just knowing the existence of such a product is not sufficient in physical application; one needs the parameters of the displacement, squeeze and rotation operators.
The time-evolution equation of $\{ \alpha (t) \}$, $\{ \zeta (t) \}$ and $\{\phi(t) \}$ is complicated and non-linear~\cite{Schumaker:1986tlu, Ma:1990llj}.
In this section, we introduce new parameters, which follow a simple linear equation.

\subsection{Hamiltonian and Heisenberg equation}\label{sec:setup}
Our Hamiltonian consists of the free and interaction (potential) parts:
\bal
& \hat{H}_0 = \hbar \sum_i \omega_i \hat{a}^\dagger_i \hat{a}_i \,, \notag \\
& \hat{V}(t) = i \hbar \sum_i \left(f_i(t) e^{-i \omega_i t} \hat{a}^\dagger_i - f^*_i(t) e^{i \omega_i t} \hat{a}_i \right) + \frac{i}{2} \hbar \sum_{i,j} \left[ g^*_{ij}(t) e^{i (\omega_i + \omega_j) t} \hat{a}_i \hat{a}_j - g_{ij}(t) e^{- i (\omega_i + \omega_j) t} \hat{a}^\dagger_i \hat{a}^\dagger_j \right]  \notag \\
& \qquad \quad + \hbar \sum_{i,j} h_{ij}(t) \hat{a}^\dagger_i \hat{a}_j \,.
\eal
Here $g(t) = g^T(t)$ and $h(t) = h^\dagger(t)$ in the matrix notation to make Hamiltonian Hermitian.
This separation is mathematically artificial in the sense that the free part can be eliminated by the redefinition of $\{h(t)\}$, but physically meaningful in the sense that $\{f(t)\}$, $\{g(t)\}$ and $\{h(t)\}$ originate from coupling to an external classical field. 
Photons via inverse Primakoff effect (from dark-matter axion) and via inverse Gertsenshtein effect (from gravitons) correspond to $\{h(t)\}=\{g(t)\}=0$ (Galuber case).
Though, for gravitons from the black-hole mergers, one needs to keep all $\{f(t)\}$, $\{g(t)\}$ and $\{h(t)\}$.

The time-evolution (unitary) operator is generically given by
\bal
&\hat{U}(t) = \hat{U}_0(t) \hat{U}_I(t) \,, \quad 
\hat{U}_0(t) = \exp\left( - i  \sum_i \omega_i t \hat{a}^\dagger_i \hat{a}_i \right) \,, \notag \\
&i\hbar \frac{d}{dt} \hat{U}_I(t) =  \hat{H}_I (t) \hat{U}_I(t)\,, \quad
\hat{H}_I (t) = \exp\left( i \sum_i \omega_i t \hat{a}^\dagger_i \hat{a}_i \right) \hat{V}(t) \exp\left( - i \sum_i \omega_i t \hat{a}^\dagger_i \hat{a}_i \right) \,,
\label{eq:evol_U}
\eal
in the interaction picture.
One can check that it satisfies
\bal
i\hbar \frac{d}{dt} \hat{U}(t) = \hat{H} (t) \hat{U}(t) \,.
\eal
For our Hamiltonian,
\bal
&\hat{U}_0(t) = \hat{P} \left( \{\omega t\} \right) \,, \notag \\
&\hat{H}_I (t) = i \hbar \sum_i \left[ f_i(t) \hat{a}^\dagger_i - f^*_i(t) \hat{a}_i \right] + \frac{i}{2} \hbar \sum_{i,j} \left[ g^*_{ij}(t) \hat{a}_i \hat{a}_j - g_{ij}(t) \hat{a}^\dagger_i \hat{a}^\dagger_j \right] + \hbar \sum_{i,j} h_{ij}(t) \hat{a}^\dagger_i \hat{a}_j \,.
\eal

Instead of directly discussing time evolution of $\hat{U}_I(t)$ (for a certain reason discussed below), we discuss time evolution (Heisenberg equation) of the following time-dependent annihilation and creation operators:
\bal
\hat{b}_i(t) = \hat{U}^\dagger_I(t) \hat{a}_i \hat{U}_I(t) \,, \quad \hat{b}^\dagger_i(t) = \hat{U}^\dagger_I(t) \hat{a}^\dagger_i \hat{U}_I(t) \,.
\eal
By using the time-evolution equation of $\hat{U}_I(t)$, one can derive the time-evolution equation of $\{\hat{b}(t)\}$:
\bal
\frac{d}{dt} \hat{b}_i(t) = \hat{U}^\dagger_I(t)  \frac{i}{\hbar} [\hat{H}_I(t), \hat{a}_i] \hat{U}_I(t) &= \hat{U}^\dagger_I(t)  \left[ f_i (t) - \sum_{j} g_{ij}(t) \hat{a}^\dagger_j - i \sum_j h(t) \hat{a}_j \right] \hat{U}_I(t) \notag \\
&= f_i(t) - \sum_{j} g_{ij}(t) \hat{b}^\dagger_j(t) - i \sum_{j} h_{ij}(t) \hat{b}_j(t) \,.
\eal
In the last equality, we have used the definition of the time-dependent annihilation (creation) operator.
Similarly, the time-evolution equation of $\{\hat{b}^\dagger(t)\}$ is given by
\bal
\frac{d}{dt} \hat{b}^\dagger_i(t) = f^*_i(t) - \sum_{j} g^*_{ij}(t) \hat{b}_j(t) + i \sum_j h^*_{ij}(t) \hat{b}^\dagger_j(t) \,.
\eal
One can rewrite the evolution equations in the matrix notation:
\bal
\frac{d}{dt}
\begin{pmatrix}
   \hat{b}(t) \\
   \hat{b}^\dagger(t)
\end{pmatrix}
= M(t)
\begin{pmatrix}
   \hat{b}(t) \\
   \hat{b}^\dagger(t)
\end{pmatrix}
+ \begin{pmatrix}
   f(t) \\
   f^*(t)
\end{pmatrix} \,, \quad
M(t) = 
\begin{pmatrix}
   - i h (t) & - g(t) \\
  - g^*(t) & i h^* (t)
\end{pmatrix}\,.
\label{eq:b-evol}
\eal
The initial condition is given at $t = t_0$ as $\hat{b}_i(t_0) = \hat{a}_i$ ($\hat{b}^\dagger_i(t_0) = \hat{a}^\dagger_i$).

\subsection{Product of the displacement, squeeze and rotation operators}\label{sec:const}
Generalized Glauber theorem states that the time-evolution operator is given by a product of the displacement, squeeze and rotation operators:
\bal
\hat{U}'_I(t) =  \hat{D}(\{\alpha(t)\}) \hat{S} (\{\zeta(t)\}) \hat{P} (\{\phi(t)\})  \,,
\eal
up to a phase.
On the other hand, it is not obvious to find time-evolution equation of $\hat{U}'_I(t)$ except for special cases discussed in the next section (see also appendix~\ref{sec:derivative}).
This is why instead we discuss time evolution of the following time-dependent annihilation and creation operators:
\bal
\hat{b}'_i(t) = \hat{U}'^\dagger_I(t) \hat{a}_i \hat{U}'_I(t) \,, \quad \hat{b}'^\dagger_i(t) = \hat{U}'^\dagger_I(t) \hat{a}^\dagger_i \hat{U}'_I(t) \,.
\eal
By using the transformation of the annihilation and reaction operators under the displacement, squeeze and rotation operators as discussed in section~\ref{sec:def}, one can find
\bal
&\hat{b}'_i(t) = \sum_j \left[ \mu^*_{ij}(t) \hat{a}_j - \nu_{ij}(t) \hat{a}^\dagger_j \right] + \alpha_i(t) \,, \quad
\hat{b}'^\dagger_i(t) = \sum_j \left[ \mu_{ij} (t) \hat{a}^\dagger_j - \nu^*_{ij}(t) \hat{a}_j \right] + \alpha^*_i(t) \,,
\eal
where in the matrix notation
\bal
&\mu(t) = \cosh \left( \left[ \zeta^*(t) \zeta(t) \right]^{1/2} \right) e^{i \phi^* (t)} \,, \notag \\
&\nu (t) = \sinh \left( \left[ \zeta(t) \zeta^*(t) \right]^{1/2} \right) \left[\zeta(t) \zeta^*(t)\right]^{-1/2} \zeta(t) e^{i \phi^* (t)} = \zeta(t) \left[\zeta^*(t) \zeta(t) \right]^{-1/2} \sinh\left( \left[ \zeta^*(t) \zeta(t) \right]^{1/2} \right) e^{i \phi^* (t)} \,,
\label{eq:new_params}
\eal
which satisfy
\bal
& \mu^T(t) \mu^*(t) - \nu^T(t) \nu^*(t) = I \,, \quad \mu^T(t) \nu(t) - \nu^T(t) \mu (t) = 0 \,, \notag \\
& \mu^*(t) \mu^T(t) - \nu(t) \nu^\dagger(t) = I \,, \quad \mu^*(t) \nu^T(t) - \nu(t) \mu^\dagger (t) = 0 \,.
\label{eq:variables-definition}
\eal

By noting the inversion relation,
\bal
&\hat{a}_i = \sum_j \left[ \mu_{ji}(t) \left( \hat{b}'_j(t) - \alpha_j(t) \right) + \nu_{ji}(t) \left( \hat{b}'^\dagger_j(t) - \alpha^*_j(t) \right) \right] \,, \notag \\
&\hat{a}^\dagger_i = \sum_j \left[ \mu^*_{ji}(t) \left( \hat{b}'^\dagger_j(t) - \alpha^*_j(t) \right) + \nu^*_{ji}(t) \left( \hat{b}'_j(t) - \alpha_j(t) \right) \right] \,,
\eal
one can derive the time-evolution equation of $\{\hat{b}'(t)\}$:
\bal
\frac{d}{dt} \hat{b}'_i(t) &= \sum_j \left( \left[ \frac{d}{dt} \mu^*_{ij}(t) \right] \hat{a}_j - \left[ \frac{d}{dt} \nu_{ij}(t) \right] \hat{a}^\dagger_j  \right) + \frac{d}{dt} \alpha_i(t) \notag \\
&= \sum_{j} \left( \left[ \frac{d}{dt} \mu^*(t) \right] \mu^T(t) - \left[ \frac{d}{dt} \nu(t) \right] \nu^\dagger (t) \right)_{ij} \hat{b}'_j(t) \notag \\
& \quad + \sum_{j} \left( \left[ \frac{d}{dt} \mu^*(t) \right] \nu^T(t) - \left[ \frac{d}{dt} \nu(t) \right] \mu^\dagger(t) \right)_{ij} \hat{b}'^\dagger_j(t)  \notag \\
& \quad + \frac{d}{dt} \alpha_i(t) - \sum_{j} \left[ \left( \left[ \frac{d}{dt} \mu^*(t) \right] \mu^T(t) - \left[ \frac{d}{dt} \nu(t) \right] \nu^\dagger (t) \right)_{ij} \alpha_j (t) \right. \notag \\
& \quad \left. - \left( \left[ \frac{d}{dt} \mu^*(t) \right] \nu^T(t) - \left[ \frac{d}{dt} \nu(t) \right] \mu^\dagger(t) \right)_{ij} \alpha^*_j(t) \right]\,.
\eal
Similarly, the time-evolution equation of $\{\hat{b}'^\dagger(t)\}$ is given by
\bal
\frac{d}{dt} \hat{b}'^\dagger_i(t) &= \sum_j \left( \left[ \frac{d}{dt} \mu_{ij}(t) \right] \hat{a}^\dagger_j - \left[ \frac{d}{dt} \nu^*_{ij}(t) \right] \hat{a}_j  \right) + \frac{d}{dt} \alpha^*_i(t) \notag \\
&= \sum_{j} \left( \left[ \frac{d}{dt} \mu(t) \right] \mu^\dagger(t) - \left[ \frac{d}{dt} \nu^*(t) \right] \nu^T (t) \right)_{ij} \hat{b}'^\dagger_j(t) \notag \\
& \quad + \sum_{j} \left( \left[ \frac{d}{dt} \mu(t) \right] \nu^\dagger(t) - \left[ \frac{d}{dt} \nu^*(t) \right] \mu^T(t) \right)_{ij} \hat{b}_j(t)  \notag \\
& \quad + \frac{d}{dt} \alpha^*_i(t) - \sum_{j} \left[ \left( \left[ \frac{d}{dt} \mu(t) \right] \mu^\dagger(t) - \left[ \frac{d}{dt} \nu^*(t) \right] \nu^T (t) \right)_{ij} \alpha^*_j (t) \right. \notag \\
& \quad \left. - \left( \left[ \frac{d}{dt} \mu(t) \right] \nu^\dagger(t) - \left[ \frac{d}{dt} \nu^*(t) \right] \mu^T(t) \right)_{ij} \alpha_j(t) \right]\,.
\eal
The initial condition is $\hat{b}'_i(t_0) = \hat{a}_i$ ($\hat{b}'^\dagger_i(t_0) = \hat{a}^\dagger_i$).

For these time-evolution equations to be the same as for $\{\hat{b}'_i(t)\}$ ($\{\hat{b}'^\dagger(t)\}$) [see \eqref{eq:b-evol}]: namely,
\bal
\frac{d}{dt}
\begin{pmatrix}
   \hat{b}'(t) \\
   \hat{b}'^\dagger(t)
\end{pmatrix}
= M(t)
\begin{pmatrix}
   \hat{b}'(t) \\
   \hat{b}'^\dagger(t)
\end{pmatrix}
+ \begin{pmatrix}
   f(t) \\
   f^*(t)
\end{pmatrix} \,,
\eal
the followings should hold:
\bal
& \left( \left[ \frac{d}{dt} \mu^*(t) \right] \mu^T(t) - \left[ \frac{d}{dt} \nu(t) \right] \nu^\dagger (t) \right)_{ij} = - i h_{ij}(t) \,, \notag \\
& \left( \left[ \frac{d}{dt} \mu(t) \right] \mu^\dagger(t) - \left[ \frac{d}{dt} \nu^*(t) \right] \nu^T (t) \right)_{ij} = i h^*_{ij}(t) \,, \notag \\
& \left( \left[ \frac{d}{dt} \mu^*(t) \right] \nu^T(t) - \left[ \frac{d}{dt} \nu(t) \right] \mu^\dagger(t) \right)_{ij} = - g_{ij}(t) \,, \notag \\
& \left( \left[ \frac{d}{dt} \mu(t) \right] \nu^\dagger(t) - \left[ \frac{d}{dt} \nu^*(t) \right] \mu^T(t) \right)_{ij} = - g^*_{ij}(t) \,, \notag \\
& \frac{d}{dt} \alpha_i(t) - \sum_{j} \left[ \left( \left[ \frac{d}{dt} \mu^*(t) \right] \mu^T(t) - \left[ \frac{d}{dt} \nu(t) \right] \nu^\dagger (t) \right)_{ij} \alpha_j (t) \right. \notag \\
& \quad \left. - \left( \left[ \frac{d}{dt} \mu^*(t) \right] \nu^T(t) - \left[ \frac{d}{dt} \nu(t) \right] \mu^\dagger(t) \right)_{ij} \alpha^*_j(t) \right] = f_i(t) \,, \notag \\
& \frac{d}{dt} \alpha^*_i(t) - \sum_{j} \left[ \left( \left[ \frac{d}{dt} \mu(t) \right] \mu^\dagger(t) - \left[ \frac{d}{dt} \nu^*(t) \right] \nu^T (t) \right)_{ij} \alpha^*_j (t) \right. \notag \\
& \quad \left. - \left( \left[ \frac{d}{dt} \mu(t) \right] \nu^\dagger(t) - \left[ \frac{d}{dt} \nu^*(t) \right] \mu^T(t) \right)_{ij} \alpha_j(t) \right] = f^*_i(t) \,.
\eal

By using \eqref{eq:variables-definition}, one can rewrite these equation as time-evolution equation of $\{ \mu (t) \}$, $\{ \nu (t) \}$ and $\{ \alpha (t) \}$:
\bal
&\frac{d}{dt}
\begin{pmatrix}
   \mu^*(t) \\
   - \nu^*(t)
\end{pmatrix}
= M(t)
\begin{pmatrix}
   \mu^*(t) \\
   - \nu^*(t)
\end{pmatrix} \,, \quad
\frac{d}{dt}
\begin{pmatrix}
   - \nu(t) \\
   \mu(t)
\end{pmatrix}
= M(t)
\begin{pmatrix}
   - \nu(t) \\
   \mu(t)
\end{pmatrix} \,, \notag \\
&\frac{d}{dt}
\begin{pmatrix}
   \alpha(t) \\
   \alpha^*(t)
\end{pmatrix}
= M(t)
\begin{pmatrix}
   \alpha(t) \\
   \alpha^*(t)
\end{pmatrix} 
+ \begin{pmatrix}
   f(t) \\
   f^*(t)
\end{pmatrix}  \,.
\label{eq:matrix-form}
\eal
The initial condition is $\mu(t_0) = I$, $\nu(t_0) = 0$ and $\alpha (t_0) = 0$.
The evolution matrix $M(t)$ satisfies
\bal
\Omega^T M(t)^T \Omega = -M(t) \,, \quad
\Omega =
\begin{pmatrix}
   0 & I \\
  - I & 0
\end{pmatrix} \,,
\eal
to preserve equation~\eqref{eq:variables-definition}.
Note that these equations are simple and linear, and actually the same as for (homogeneous part of) $\{\hat{b}'_i(t)\}$ ($\{\hat{b}'^\dagger(t)\}$) [see \eqref{eq:b-evol}].

In summary, since they satisfy the same evolution equation with the same initial condition, $\{\hat{b}'(t)\} = \{\hat{b}(t)\}$ ($\{\hat{b}'^\dagger(t)\} = \{\hat{b}^\dagger(t)\}$).
Therefore, $\hat{U}'_I(t) = \hat{U}_I(t)$ up to a phase.
We remark that one can interchange the order of the displacement, squeeze and rotation operators with a proper change of the parameters.
Though we do not exhaust all the cases, an (possibly most non-trivial) example is 
yet another form of $\hat{U}'_I(t) = \hat{S} (\{\zeta(t)\}) \hat{D}(\{\tilde{\alpha}(t)\}) \hat{P} (\{\phi(t)\})$ with $\hat{S}^\dagger (\{\zeta(t)\}) \hat{D}(\{\alpha(t)\}) \hat{S} (\{\zeta(t)\}) = \hat{D}(\{\tilde{\alpha}(t)\})$.
By using the transformation of the annihilation and creation operators under the squeeze operator, one finds
\bal
\tilde{\alpha}_i(t) = \sum_j \left[ \mu_{ji}(t) \alpha_j(t) + \nu_{ji}(t) \alpha^*_j(t) \right] \,, \quad \tilde{\alpha}^*_i(t) = \sum_j \left[ \mu^*_{ji}(t) \alpha^*_j(t) + \nu^*_{ji}(t) \alpha_j(t) \right] \,,
\eal
where
\bal
& \mu(t) = \cosh \left( [\zeta^*(t) \zeta(t)]^{1/2} \right) \,, \notag \\
& \nu(t) = \sinh \left( [\zeta(t) \zeta^*(t)]^{1/2} \right) [\zeta(t) \zeta^*(t)]^{-1/2} \zeta(t) = \zeta(t) [\zeta^*(t) \zeta(t)]^{-1/2} \sinh \left( [\zeta^*(t) \zeta(t)]^{1/2} \right) \,,
\eal
satisfy
\bal
\mu(t) = \mu^\dagger(t) \,, \quad \nu(t) = \nu^T(t) \,, \quad \mu^2(t) - \nu^*(t) \nu(t) = I \,, \quad \mu^*(t) \nu(t) - \nu(t) \mu(t) = 0 \,.
\eal
Physically speaking, it means that photons, produced via linear and quadratic couplings to classical sources, follow a squeezed coherent state.

\subsection{Special cases}
We discuss two special cases where one can explicitly show that $\hat{U}'_I(t)$ solves the time-evolution equation \eqref{eq:evol_U}.
\subsubsection{Dark-matter axion detection: $\{h(t)\}=\{g(t)\}=0$} \label{sec:linear}
This is the Glauber case. From \eqref{eq:new_params} and \eqref{eq:variables-definition}, one finds that $\mu_{ij}(t) = \delta_{ij}$ and $\nu_{ij}(t) = 0$, corresponding to $\zeta_{ij}(t) = \phi_{ij}(t) =0$, and
\bal
\frac{d}{dt} \alpha_i(t) = f_i(t) \,.
\eal
Therefore,
\bal
\hat{U}'_I(t) = \bigotimes_{i} \hat{D}(\alpha_i(t))  \,.
\eal
The time derivative of the displacement operator is given by (see appendix~\ref{sec:derivative})
\bal
\frac{d}{dt} \hat{D} (\alpha_i(t)) &= \left( \hat{a}^\dagger_i \frac{d}{dt} \alpha_i(t) - \hat{a}_i \frac{d}{dt} \alpha^*_i(t) + \frac{1}{2} \left[ \alpha_i(t) \frac{d}{dt} \alpha^*_i(t) - \alpha^*_i(t) \frac{d}{dt} \alpha_i(t) \right] \right) \hat{D} (\alpha_i(t)) \notag \\
&= \left( f_i(t) \hat{a}^\dagger_i - f^*_i(t) \hat{a}_i + \frac{1}{2} \left[ f^*_i(t) \alpha_i(t) - f_i(t) \alpha^*_i(t) \right] \right) \hat{D} (\alpha_i(t)) \,.
\eal
By combining it with the pure phase $\beta (t)$, one obtains
\bal
i \hbar \frac{d}{dt} \left[ e^{i \beta_i(t)}\hat{D} (\alpha_i(t)) \right] &= i \hbar \left( f_i(t) \hat{a}^\dagger_i - f^*_i(t) \hat{a}_i + \frac{1}{2} \left[ f^*_i(t) \alpha_i(t) - f_i(t) \alpha^*_i(t) \right] + i \frac{d}{dt} \beta_i(t) \right) \notag \\
&\quad \times \left[ e^{i \beta_i(t)}\hat{D} (\alpha_i(t)) \right] \,.
\eal
This solves the time-evolution equation \eqref{eq:evol_U} with $h_{ij}(t)=g_{ij}(t)=0$, when one chooses
\bal
& \frac{d}{dt} \beta_i(t) = i \frac{1}{2} \sum_i \left[ f^*_i(t) \alpha_i(t) - f_i(t) \alpha^*_i(t) \right] \,, \notag \\
& \beta_i(t) = i \frac{1}{2} \int_0^t dt_1 \int_0^{t_1} dt_2  \sum_i \left[ f^*_i(t_1) f_i(t_2) - f_i(t_1) f^*_i(t_2) \right] \,.
\eal

\subsubsection{$\{h(t)\}=\{f(t)\}=0$ and ``real'' diagonal $\{g(t)\}$}\label{sec:realg}
This is the case of parametric amplifier.
We consider $g_{ij} (t)  = \delta_{ij} g_{i} (t) e^{i \gamma_i}$ with real $g_{i} (t)$ and constant $\gamma_i$.
From \eqref{eq:matrix-form}, one finds that $\mu_{ij}(t) = \delta_{ij} \cosh(r_i(t))$, $\nu_{ij}(t) = \delta_{ij} \sinh(r_i(t)) e^{i \gamma_i} $ and $\alpha_i(t) = 0$, 
\bal
\frac{d}{dt} r_i(t) = g_i(t) \,.
\eal
From \eqref{eq:new_params}, they correspond to $\zeta_{ij}(t) = \delta_{ij} r_i(t) e^{i \gamma_i} $, $\phi_{ij}(t)=0$ and $\alpha_i(t)=0$.
Therefore,
\bal
\hat{U}'_I(t) = \bigotimes_i \hat{S}(r_i(t)e^{i \gamma_i})  \,.
\eal
The time derivative of the squeeze operator is given by (see appendix~\ref{sec:derivative})
\bal
\frac{d}{dt} \hat{S} (r_i(t)e^{i \gamma_i}) &= \frac{1}{4} \left[ \left( r_i(t) e^{-i \gamma_i} \hat{a}^2_i - r_i(t) e^{i \gamma_i} \hat{a}^{\dagger 2}_i \right) \frac{d}{dt} \ln \left( r^2_i(t) \right) \right] \hat{S} (r_i(t)e^{i \gamma_i}) \notag \\
&= \frac{1}{2} \left[  g_i(t) e^{-i \gamma_i} \hat{a}^2_i - g_i(t) e^{i \gamma_i} \hat{a}^{\dagger 2}_i \right] \hat{S} (r_i(t)e^{i \gamma_i}) \,.
\eal
This solves the time-evolution equation \eqref{eq:evol_U} with $h_{ij}(t)=f_i(t)=0$ and $g_{ij} (t)= \delta_{ij} g_{i} (t) e^{i \gamma_i}$ with real $g_{i} (t)$ and constant $\gamma_i$.

\subsection{Comparison with naive computation}\label{sec:comparison}

The time-evolution equation \eqref{eq:evol_U} is generically ``solved'' by Dyson series
\bal
\hat{U}''_I(t) = T\left\{ \exp\left[ - \frac{i}{\hbar} \int^t_{t_0} dt' \hat{H}_I (t') \right] \right\} \,. 
\label{eq:interaction-evolution}
\eal
We illustrate the role of time-ordering $T$,
\bal
T\left\{ \hat{\cal O}(t_1) \hat{\cal O}(t_2)  \right\} = \theta(t_1-t_2) \hat{\cal O}(t_1) \hat{\cal O}(t_2) + \theta(t_2-t_1)\hat{\cal O}(t_2) \hat{\cal O}(t_1) \,,
\eal
by considering the Dyson series up to the 2nd order:
\bal
\hat{U}''_I(t) &\approx T \left\{ 1  - \frac{i}{\hbar} \int^t_{t_0} dt_1 \hat{H}_I (t_1) - \frac{1}{2 \hbar^2} \int^t_{t_0} dt_1  \hat{H}_I (t_1) \int^t_{t_0} dt_2 \hat{H}_I (t_2) \right\} \notag \\
& = 1 - \frac{i}{\hbar} \int^t_{t_0} dt_1 \hat{H}_I (t_1) - \frac{1}{2 \hbar^2} \int^t_{t_0} dt_1 \int^t_{t_0} dt_2 T \left\{\hat{H}_I (t_1)  \hat{H}_I (t_2) \right\} \,.
\eal
Let us take the derivative with respect to $t$:
\bal
i \hbar \frac{d}{dt} \hat{U}''_I(t) &\approx  \hat{H}_I (t) - \frac{i}{2 \hbar} \left[ \int^t_{t_0} dt_2 T \left\{\hat{H}_I (t)  \hat{H}_I (t_2) \right\} + \int^t_{t_0} dt_1 T \left\{\hat{H}_I (t_1)  \hat{H}_I (t) \right\} \right] \notag \\
& = \hat{H}_I (t) - \frac{i}{\hbar} \hat{H}_I (t) \int^t_{t_0} dt_1 \hat{H}_I (t_1) 
\approx \hat{H}_I (t) \hat{U}''_I(t) \,.
\eal
In the second equality, we have used the property of time ordering; without it, one cannot take $\hat{H}_I (t)$ to the left most and thus the Dyson series does not solve the time-evolution equation \eqref{eq:evol_U}, unless $\hat{H}_I (t)$ commutes with $\hat{H}_I (t')$ for any $t_0 < t' < t$.
Actually, Dyson series is defined only for the series expression of the exponential (this is why it is a series).

\begin{figure}[H]
    \centering
    \begin{minipage}[b]{0.49\columnwidth}
    \centering
    \includegraphics[width=0.99\linewidth]{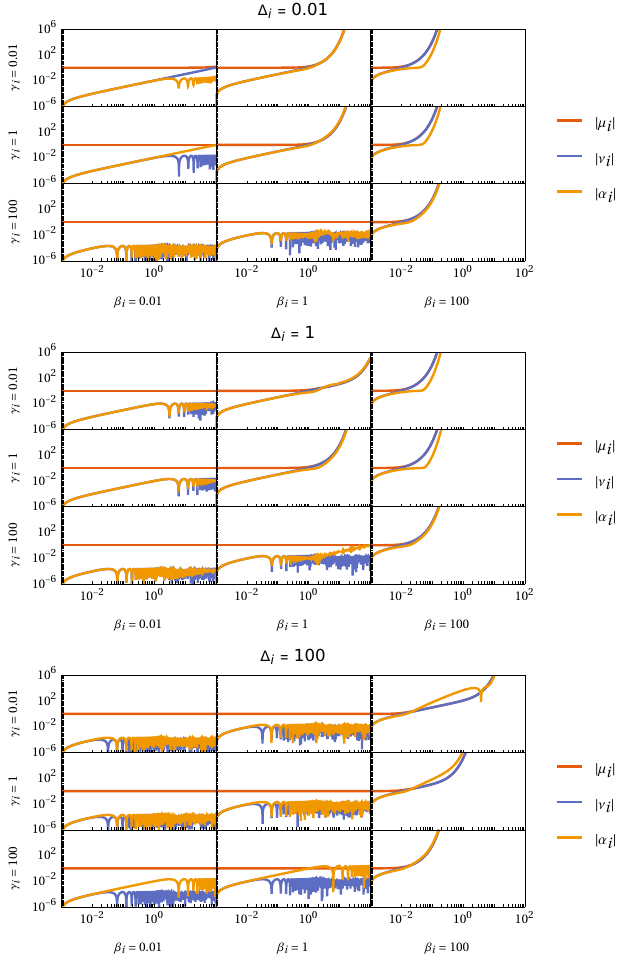}
    \end{minipage}
    \begin{minipage}[b]{0.49\columnwidth}
    \centering
    \includegraphics[width=.99\linewidth]{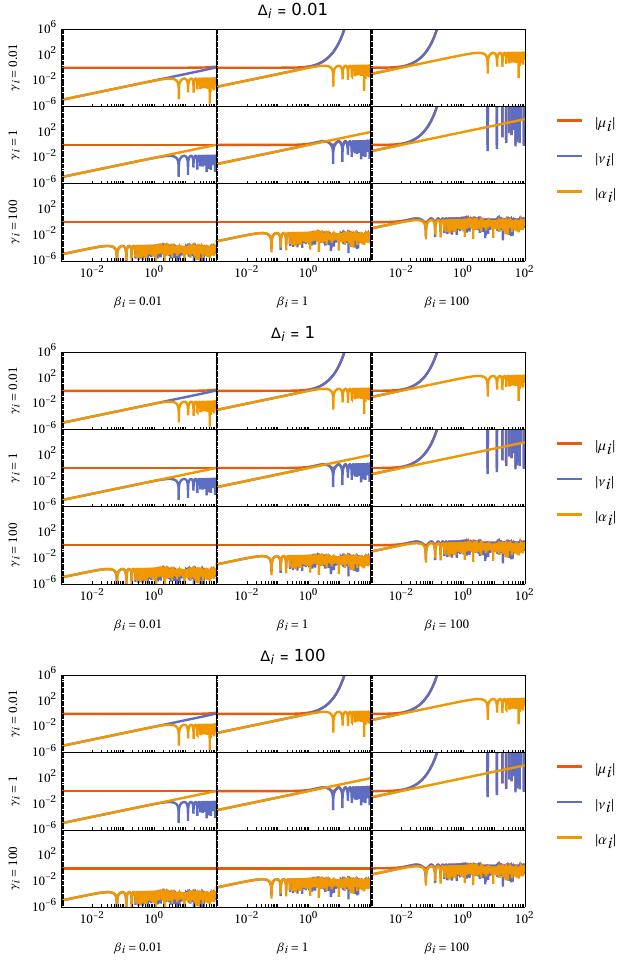}
    \end{minipage}
    \caption{Comparison between the actual (left) and naive (right) time evolution of $|\mu_i (t)|$, $|\nu_i (t)|$ and $|\alpha_i (t)|$. Here we take values of $\beta_i, \gamma_i, \Delta_i = 0.01, 1, 100$.}
    \label{fig:deltan}
\end{figure}

On the other hand, it is a common mistake computing it as if we could evaluate the exponent first and then take time-ordering.
This naive approach results in
\bal
\tilde{\alpha}_i (t) = \int_{t_0}^{t} dt_1 f_i(t_1)  \,, \quad \tilde{\zeta}_{ij} (t) = \int_{t_0}^{t} dt_1 g_{ij}(t_1) \,, \quad \tilde{\phi}_{ij} (t) = \int_{t_0}^{t} dt_1 h_{ij}(t_1) \,.
\eal
in our case.
We illustrate if and when there are sizable differences in the following.
To be concrete, we consider a case with no mode mixings: $g_{ij} (t) = \delta_{ij} g_i (t)$ and $h_{ij} (t)= \delta_{ij} h_i (t)$.
We focus on a single mode and thus omit mode indices.
More specifically, we consider the following functional forms: 
\bal
f_i(t) = \beta_i e^{-i (\gamma_i - \omega_i) t} \,, \quad g_i(t) = \beta_i e^{-i \gamma_i t} \,, \quad h_i(t) = \Delta_i \,,
\eal
where $\beta_i$, $\gamma_i$, and $\Delta_i$ are real constants.
By performing the integral analytically, one finds the naive values
\bal
& \tilde{\alpha}_i(t) =\frac{i \beta_i  z\left(t;\gamma_i -\omega_i\right)}{\gamma_i -\omega_i } \,, \quad \tilde{\mu}_i(t)=e^{i \Delta_i  t} \cosh \left(\left| \frac{\beta_i z(t; \gamma_i)}{\gamma_i }\right| \right) \notag \\
& \tilde{\nu}_i(t)=i e^{i \Delta_i  t} \frac{z(t; \gamma_i)}{|z(t; \gamma_i)|} \sinh \left( \left| \frac{\beta_i z(t; \gamma_i)}{\gamma_i }\right| \right) \,,
\eal
where $z(t; x)=-1+e^{-i x t}$ is a complex function parametrized by a real parameter $x$. 
Here we took $t_0 = 0$. 
For actual values, we solve the differential equation \eqref{eq:matrix-form} in $\omega_i = 1$ units (see appendix~\ref{sec:num_conv} for numerical convergence check).
The comparison is presented in figure \ref{fig:deltan}.

First, the naive and actual results are very similar to each other for small $\beta_i$ and small $\Delta_i$ (for arbitrary $\gamma_i$).
This is because with such a parameter, we need to keep only up to the 1st order in the Dyson series, for which time-ordering does not matter.
On the other hand, the two results differ significantly for larger values of $\beta_i$ or $\Delta_i$.
Second, note that naive (right) results do not change for different $\Delta_i$.
This is because $\Delta_i$ appears only in the phase of $\mu_i (t)$ and $\nu_i (t)$, while we plot $|\mu_i (t)|$ and $|\nu_i (t)|$.
On the other hand, actual computation (left) exhibits the $\Delta_i$ dependence.
This is because actual impact of $\Delta_i$ is shifting $\omega_i \to \omega_i + \Delta_i$.
This is why the resonant behavior (linear growth with time) of $|\alpha_i(t)|$ for $\gamma_i = 1$ in the naive computation disappears in the actual computation.

\section{Application for mix states}\label{sec:app}
The time-evolution of a density matrix is given by
\bal
\hat{\rho}(t) = \hat{U}(t) \hat{\rho}(t_0) \hat{U}^\dagger(t) = \hat{P}(\omega t) \hat{U}'(t) \hat{\rho}(t_0) \hat{U}'^\dagger(t) \hat{P}^\dagger(\omega t) \,.
\eal
We stress again that the time-dependent phase factor ignored in $\hat{U}'(t)$ does not matter at all for the density matrix.
By using the transformations of the annihilation and creation operators under the displacement, squeeze and rotation operators, one can rewrite expectation value of the annihilation (creation) operator in any functional form $F(\hat{a}, \hat{a}^\dagger)$ as:
\bal
\langle F(\{\hat{a}\}, \{\hat{a}^\dagger\}) \rangle (t) &= {\rm Tr} \left( F(\{\hat{a}\}, \{\hat{a}^\dagger\}) \hat{\rho}(t) \right)
=  {\rm Tr} \left( \hat{U}'^\dagger_I(t) \hat{P}^\dagger(\{\omega t\}) F(\{\hat{a}\}, \{\hat{a}^\dagger\}) \hat{P}(\{\omega t\}) \hat{U}'_I(t) \hat{\rho}(t_0) \right) \notag \\
&= {\rm Tr} \left( F\left(\{\hat{c} (t)\}, \{\hat{c}^\dagger(t)\}\right) \hat{\rho}(t_0) \right) \,,
\eal
where
\bal
& \hat{c}_i (t) =  \hat{U}'^\dagger_I(t) \hat{P}^\dagger(\{\omega t\}) \hat{a}_i \hat{P}(\{\omega t\}) \hat{U}'_I(t)
= e^{- i \omega_i t} \left( \sum_j \left[ \mu^*_{ij}(t) \hat{a}_j - \nu_{ij}(t) \hat{a}^\dagger_j \right] + \alpha_i(t) \right) \,, \notag \\
& \hat{c}^\dagger_i (t) =  \hat{U}'^\dagger_I(t) \hat{P}^\dagger(\{\omega t\}) \hat{a}^\dagger_i \hat{P}(\{\omega t\}) \hat{U}'_I(t)
= e^{i \omega_i t} \left( \sum_j \left[ \mu_{ij} (t) \hat{a}^\dagger_j - \nu^*_{ij}(t) \hat{a}_j \right] + \alpha^*_i(t) \right) \,.
\eal

We explicitly compute cases where $F$ contains only up to the 2nd order of the annihilation (creation) operators, which would be most relevant to evaluate the expectation value of a field and fluctuation around it.
We assume that $F$ is normal-ordered and $\hat{\rho}$ is normalized to be unity without loss of generality.
We obtain
\bal
& \langle \hat{a}_i \rangle (t) = e^{-i \omega_i t} \left( \sum_j \left[ \mu^*_{ij}(t) \langle \hat{a}_j \rangle (t_0) - \nu_{ij}(t) \langle \hat{a}^\dagger_j \rangle (t_0) + \alpha_i(t) \right] \right) \,, \notag \\
& \langle \hat{a}^\dagger_i \rangle (t) = e^{i \omega_i t} \left( \sum_j \left[ \mu_{ij}(t) \langle \hat{a}^\dagger_j \rangle (t_0) - \nu^*_{ij}(t) \langle \hat{a}_j \rangle (t_0) + \alpha^*_i(t) \right] \right) \,, \notag \\
& \langle \hat{a}_i \hat{a}_j \rangle (t) = e^{-i (\omega_i + \omega_j) t} \left( \sum_{k,l} \left[ \mu^*_{ik}(t) \mu^*_{jl}(t) \langle \hat{a}_k \hat{a}_l \rangle (t_0) + \nu_{ik}(t) \nu_{jl}(t) \langle \hat{a}^\dagger_k \hat{a}^\dagger_l \rangle (t_0) - \mu^*_{ik}(t) \nu_{jl}(t) \langle \hat{a}_k \hat{a}^\dagger_l \rangle  \right. \right. \notag \\
& \qquad \qquad \left. - \nu_{ik}(t) \mu^*_{jl}(t) \langle \hat{a}^\dagger_k \hat{a}_l \rangle (t_0) \right] + \sum_{k} \left[ \left( \mu^*_{ik}(t) \alpha_j(t) + \alpha_i(t) \mu^*_{jk}(t) \right) \langle \hat{a}_k \rangle (t_0) \right. \notag \\
& \qquad \qquad \left. \left. - \left( \nu_{ik}(t) \alpha_j(t) + \alpha_i(t) \nu_{jk}(t) \right) \langle \hat{a}^\dagger_k \rangle (t_0) \right] + \alpha_i(t) \alpha_j(t) \right) \,, \notag \\
& \langle \hat{a}^\dagger_i \hat{a}^\dagger_j \rangle (t) = e^{i (\omega_i + \omega_j) t} \left( \sum_{k,l} \left[ \mu_{ik}(t) \mu_{jl}(t) \langle \hat{a}^\dagger_k \hat{a}^\dagger_l \rangle (t_0) + \nu^*_{ik}(t) \nu^*_{jl}(t) \langle \hat{a} \hat{a}_l \rangle (t_0) - \mu_{ik}(t) \nu^*_{jl}(t) \langle \hat{a}^\dagger_k \hat{a}_l \rangle  \right. \right. \notag \\
& \qquad \qquad \left. - \nu_{ik}(t) \mu^*_{jl}(t) \langle \hat{a}_k \hat{a}^\dagger_l \rangle (t_0) \right] + \sum_k \left[ \left( \mu^*_{ik}(t) \alpha_j(t) + \alpha^*_i(t) \mu_{jk}(t) \right) \langle \hat{a}^\dagger_k \rangle (t_0) \right. \notag \\
& \qquad \qquad \left. \left. - \left( \nu^*_{ik}(t) \alpha^*_j(t) + \alpha^*_i(t) \nu^*_{jk}(t) \right) \langle \hat{a}_k \rangle (t_0) \right] + \alpha^*_i(t) \alpha^*_j(t) \right) \,, \notag \\
& \langle \hat{a}^\dagger_i \hat{a}_j \rangle (t) = e^{i (\omega_i - \omega_j) t} \left( \sum_{k,l} \left[ \mu_{ik}(t) \mu^*_{jl}(t) \langle \hat{a}^\dagger_k \hat{a}_l \rangle (t_0) + \nu^*_{ik}(t) \nu_{jl}(t) \langle \hat{a}_k \hat{a}^\dagger_l \rangle (t_0) - \mu_{ik}(t) \nu_{jl}(t) \langle \hat{a}^\dagger_k \hat{a}^\dagger_l \rangle (t_0) \right. \right. \notag \\
& \qquad \qquad \left. - \mu^*_{ik}(t) \nu^*_{jl}(t) \langle \hat{a}_k \hat{a}_l \rangle (t_0) \right] + \sum_k \left[ \left( \mu_{ik}(t) \alpha_j(t) - \alpha^*_i(t) \nu_{jk}(t) \right) \langle \hat{a}^\dagger_l \rangle (t_0) \right.  \notag \\
& \qquad \qquad \left. \left. + \left( \alpha^*_i(t) \mu^*_{jk}(t) - \nu^*_{ik}(t) \alpha_j(t) \right) \langle \hat{a}_k \rangle (t_0) \right] + \alpha^*_i(t) \alpha_j(t) \right) \,.
\eal

In particular, the results for the initial thermal state,
\bal
\hat{\rho}(t_0) = \prod_{i} \left[1 - \exp\left(- \frac{\hbar \omega_i}{k_{\rm B} T} \right) \right] \exp\left(- \frac{\hbar \omega_i }{k_{\rm B} T} \hat{a}^\dagger_{i} \hat{a}_{i} \right) \,,
\eal
would be useful.
The initial thermal state satisfy
\bal
&\langle \prod_n \hat{a}_{i_n} \rangle (t_0) = \langle \prod_n \hat{a}^\dagger_{i_n} \rangle (t_0) = 0 \,, \notag \\ 
&\langle \prod_m \hat{a}^\dagger_{i_m} \prod_n \hat{a}_{i_n} \rangle (t_0) = \text{all possible full pairings.}
\label{eq:therm}
\eal
The last one follows from the Wick theorem. Each pairing of annihilation and creation operators like $\hat{a}^\dagger_i \hat{a}_j$ is replaced by its expectation value:
\bal
\delta_{ij} \frac{1}{\exp\left(\frac{\hbar \omega_i}{k_{\rm B} T} \right) - 1} = \delta_{ij} n_{i}(T) \,.
\eal
Full pairings mean that every annihilation (creation) operator is paired with a creation (annihilation) operator; otherwise, it does not contribute. Therefore, for $m \neq n$, the expectation value is $0$. 
By using these properties, we obtain
\bal
& \langle \hat{a}_i \rangle (t) = e^{-i \omega_i t} \alpha_i(t) \,, \quad \langle \hat{a}^\dagger_i \rangle (t) = e^{i \omega _it} \alpha^*_i(t) \,, \notag \\
& \langle \hat{a}_i \hat{a}_j \rangle (t) = e^{-i (\omega_i + \omega_j) t} \left( - \sum_{k} \left[ \mu^*_{ik}(t) \nu_{jk}(t) \left(n_k(T) + 1 \right) + \nu_{ik}(t) \mu^*_{jk}(t) n_k(T) \right] + \alpha_i(t) \alpha_j(t) \right) \,, \notag \\
& \langle \hat{a}^\dagger_i \hat{a}^\dagger_j \rangle (t) = e^{i (\omega_i + \omega_j) t} \left( - \sum_{k} \left[ \mu_{ik}(t) \nu^*_{jk}(t) n_k(T) + \nu^*_{ik}(t) \mu_{jk}(t) \left(n_k(T) + 1 \right) \right] + \alpha^*_i(t) \alpha^*_j(t) \right) \,, \notag \\
& \langle \hat{a}^\dagger_i \hat{a}_j \rangle (t) = e^{i (\omega_i - \omega_j) t} \left( \sum_k \left[ \mu_{ik}(t) \mu^*_{jk}(t) n_k(T) + \nu^*_{ik}(t) \nu_{jk}(t) \left(n_k(T) + 1 \right) \right] + \alpha^*_i(t) \alpha_j(t) \right)\,.
\label{eq:gen_exp}
\eal

\subsection{Dark-matter axion detection: $\{h(t)\}=\{g(t)\}=0$}
As shown in section~\ref{sec:linear}, in the Glauber case, the evolution operator is just given by the displacement operator.
Therefore the density matrix is given by the thermal coherent state. It satisfies
\bal
& \langle \prod_n \hat{a}_{i_n} \rangle (t) = \prod_n e^{- i \omega_{i_n} t} \alpha_{i_n}(t) \,, \quad \langle \prod_n \hat{a}^\dagger_{i_n} \rangle (t) = \prod_n e^{i \omega_{i_n} t} \alpha^*_{i_n}(t) \,, \notag \\
& \langle \prod_m \hat{a}^\dagger_{i_m} \prod_n \hat{a}_{i_n} \rangle (t) = \text{all possible pairings and replacements.}
\label{eq:therm_coh}
\eal
Note that unlike thermal state \eqref{eq:therm}, the last line does not require full pairings; a number of pairings is arbitrary (including no pairing). Replacements mean that the paired annihilation and creation operators like $\hat{a}^\dagger_i \hat{a}_j$ are replaced by $\delta_{ij} n_i(T)$ as for thermal state, and then the remaining unpaired annihilation (creation) operators are replaced by $e^{- i \omega_i t} \alpha_i(t)$ [$e^{i \omega_i t} \alpha^*_i(t)$].
For example, one obtains
\bal
& \langle \hat{a}_i \rangle (t) = e^{-i \omega_i t} \alpha_i(t) \,, \quad \langle \hat{a}^\dagger_i \rangle (t) = e^{i \omega _it} \alpha^*_i(t) \,, \notag \\
& \langle \hat{a}_i \hat{a}_j \rangle (t) = e^{-i (\omega_i + \omega_j) t} \alpha_i(t) \alpha_j(t) \,, \quad \langle \hat{a}^\dagger_i \hat{a}^\dagger_j \rangle (t) = e^{i (\omega_i + \omega_j) t} \alpha^*_i(t) \alpha^*_j(t) \notag \\
& \langle \hat{a}^\dagger_i \hat{a}_j \rangle (t) = \delta_{ij} n_i(T) + e^{i (\omega_i - \omega_j) t} \alpha^*_i(t) \alpha_j(t) \,.
\eal
One can see that they agree with the general results \eqref{eq:gen_exp} with $\mu_{ij}(t) = \delta_{ij}$ and $\nu_{ij}(t) = 0$.
Furthermore,
\bal
\langle \hat{a}^\dagger_i \hat{a}_j \hat{a}^\dagger_k \hat{a}_l \rangle (t) &= \langle \hat{a}^\dagger_i \hat{a}^\dagger_k \hat{a}_j \hat{a}_l \rangle (t) + \delta_{jk} \langle \hat{a}^\dagger_i \hat{a}_l \rangle (t) \notag \\
&= \delta_{ij} \delta_{kl} n_i(T) n_j(T) + \delta_{il} \delta_{jk} n_i(T) \left( n_j(T) + 1 \right) + \delta_{ij} n_i(T) e^{i (\omega_k - \omega_l) t} \alpha^*_k(t) \alpha_l(t) \notag \\
& \quad + \delta_{il} n_i(T) e^{i (\omega_k - \omega_j) t} \alpha^*_k(t) \alpha_j(t) + \delta_{jk} (n_j(T)+1) e^{i (\omega_i - \omega_l) t} \alpha^*_i(t) \alpha_l(t) \notag \\
& \quad + \delta_{kl} n_k(T) e^{i (\omega_i - \omega_j) t} \alpha^*_i(t) \alpha_j(t) + e^{i (\omega_i + \omega_k - \omega_j - \omega_j) t} \alpha^*_i(t) \alpha^*_k(t) \alpha_j(t) \alpha_l(t)
\eal
would be useful to discuss variance of the number operators.
We remark that while the expectation value of the number operator is separable into thermal and quantum one, the variance has a mixed (thermal times quantum) contribution as seen in the second line of the right-most side.

There is yet another description of thermal coherent state.
First, note that the coherent-state representation of the thermal state
\bal
\hat{\rho}(t_0) =  \bigotimes_i \left[ \sum_{n_i} \frac{1}{n_i(T) + 1} \left( \frac{n_i(T)}{n_i(T) + 1} \right)^{n_i} |n_i \rangle \langle n_i| \right] = \bigotimes_i \left[\int \frac{d^2 \beta_i}{\pi} \frac{1}{n_i(T)} \exp\left(- \frac{\beta^*_i \beta_i}{n_i(T)}\right) |\beta_i \rangle \langle \beta_i| \right]\,.
\eal
One can see the second equality by rewriting a coherent state in terms of number states \eqref{eq:coh_state}.
Second, note that
\bal
\hat{D}(\alpha_i) \hat{D}(\beta_i) &= \exp \left[ (\alpha_i + \beta_i) a^\dagger_i - (\alpha^*_i + \beta^*_i) a_i + \frac{1}{2} \left( \alpha_i \beta^*_i - \alpha^*_i \beta_i \right) \right] \notag \\
&= \exp\left[ \frac{1}{2} \left( \alpha_i \beta^*_i - \alpha^*_i \beta_i \right) \right] \hat{D}(\alpha_i+\beta_i) \,.
\eal
In the first equality, we have used (weaker) BCH theorem and kept the phase factor, though it does not impact any observables.
By using these two, one obtains
\bal
\hat{\rho}(t) &= \bigotimes_i \left[\int \frac{d^2 \beta_i}{\pi} \frac{1}{n_i(T)} \exp\left(- \frac{\beta^*_i \beta_i}{n_i(T)}\right) \hat{D}\left( e^{-i \omega_i t}\alpha_i(t)+\beta_i \right)  |0_i \rangle \langle 0_i| \hat{D}^\dagger\left( e^{-i \omega_i t}\alpha_i(t)+\beta_i \right) \right] \notag \\
&= \bigotimes_i \left[\int \frac{d^2 \beta_i}{\pi} \frac{1}{n_i(T)} \exp\left(- \frac{[\beta_i - e^{-i \omega_i t} \alpha_i(t)]^* [\beta_i - e^{-i \omega_i t} \alpha_i(t)]}{n_i(T)}\right) |\beta_i \rangle \langle \beta_i| \right] \,.
\eal
One can evaluate the expectation value of any normal-ordered product of annihilation and creation operators as follows:
\bal
\langle \hat{a}^{\dagger m}_i \hat{a}_i^n \rangle = \int \frac{d^2 \beta_i}{\pi} \beta^{* m}_i \beta_i^{n} \frac{1}{n_i(T)} \exp\left(- \frac{[\beta_i - e^{-i \omega_i t} \alpha_i(t)]^* [\beta_i - e^{-i \omega_i t} \alpha_i(t)]}{n_i(T)}\right) \,.
\eal
Therefore one may interpret
\bal
p(\beta_i) = \frac{1}{\pi} \frac{1}{n_i(T)} \exp\left(- \frac{[\beta_i - e^{-i \omega_i t} \alpha_i(t)]^* [\beta_i - e^{-i \omega_i t} \alpha_i(t)]}{n_i(T)}\right) 
\eal
as a ``classical'' probability density.


\section{Conclusions}\label{sec:conclusions}
We revisited the generalized Glauber theorem upon recent interests in particle physics  (dark-matter axion and graviton detection).
It states that the time-evolution operator is given by a product of displacement, squeeze and rotation operators, when Hamiltonian consists of up to the 2nd-order terms in creation and annihilation operators interacting classical fields.
To determine parameters of the displacement, squeeze and rotation operators, we derive differential equations they follow.
Therefore, analytical or simple numerical analyses are sufficient to describe the quantum state of secondary particles generated from a classical primary fields.
The important application is to squeezed gravitons from the black-hole mergers.
We also compared this formalism based on generalized Glauber theorem with Dyson series.
We demonstrated the importance of time ordering in Dyson series.

The theoretical framework of the axion detection via microwave photons is described as a displacement operator and is a special case of this theorem with up to the 1st-order terms.
The theorem has been applied to thermal state of microwave photons not only the vacuum state, providing the theoretical background of axion experiments.
On-going axion searches heuristically estimate the noise level without considering and sometimes naively project it to the sensitivity of future quantum-enhanced search.
In the truly quantum regime, one needs to develop a reliable Monte Carlo simulation from the quantum statistics of the axion experiments.
Even in the state-of-the-art dilution refrigerators, microwave photons inside the axion detectors are thermalized to some finite temperature.
The thermal coherent state, revisited in this paper, provides a quantum model of axion detectors more precise than a simple coherent state.
Despite the simple mean values, i.e. quantum + thermal, the variance and thus standard deviation behave non-trivially and should be accounted in future simulation codes by introducing the probability density shown in this paper.

\appendix

\section{Useful identities} \label{sec:identities}
BCH lemma concerns a similar transformation of $\hat{Y}$ by $e^{\hat{X}}$:
\bal
e^{\hat{X}} \hat{Y} e^{- \hat{X}} = \sum_{n=0}^{\infty} \frac{1}{n!} [\hat{X}, \hat{Y}]_{n} \,,
\eal
where $[\hat{X}, \hat{Y}]_{n}$ is a recursive commutator as
\bal
[\hat{X}, \hat{Y}]_{n} = [\hat{X}, [\hat{X}, \hat{Y}]_{n-1}] \,,
\eal
with $[\hat{X}, \hat{Y}]_{n=0} = \hat{Y}$.
One can prove it by introducing
\bal
\hat{Y}(\alpha) = e^{\alpha \hat{X}} \hat{Y} e^{- \alpha \hat{X}} \,.
\eal
$\hat{Y}(\alpha=1)$ is the left-hand side of the lemma with the initial condition $\hat{Y}(\alpha=0) = \hat{Y}$.
By taking the derivative with respect to $\alpha$, one finds the ordinary differential equation:
\bal
\frac{d}{d\alpha} \hat{Y}(\alpha) = e^{\alpha \hat{X}} [\hat{X}, \hat{Y}] e^{- \alpha \hat{X}} = [\hat{X}, \hat{Y}(\alpha)] \,.
\eal
One can also introduce
\bal
\hat{Y}'(\alpha) = \sum_{n=0}^{\infty} \frac{\alpha^n}{n!} [\hat{X}, \hat{Y}]_{n} \,.
\eal
$\hat{Y}'(\alpha=1)$ is the right-hand side of the lemma with the initial condition $\hat{Y}'(\alpha=0) = \hat{Y}$.
By taking the derivative with respect to $\alpha$, one finds the ordinary differential equation:
\bal
\frac{d}{d\alpha} \hat{Y}'(\alpha) = [\hat{X}, \sum_{n=1}^{\infty} \frac{\alpha^{n-1}}{(n-1)!} [\hat{X}, \hat{Y}]_{n-1}] = [\hat{X},  \hat{Y}'(\alpha)] \,.
\eal
Therefore, both $\hat{Y}(\alpha)$ and $\hat{Y}'(\alpha)$ satisfy the same ordinary differential equation with the same initial condition.
It follows that $\hat{Y}(\alpha) = \hat{Y}'(\alpha)$ including $\alpha =1$.

(Weaker) BCH theorem concerns a product of
\bal
e^{\hat{X}} e^{\hat{Y}} = e^{\hat{Z}} \,, \quad \hat{Z} = \hat{X} + \hat{Y} + \frac{1}{2} [\hat{X}, \hat{Y}] \,,
\eal
when $[\hat{X}, \hat{Y}]$ commutes with $\hat{X}$ and $\hat{Y}$.
One can prove it by introducing
\bal
{\hat W}(\alpha) = e^{\alpha \hat{X}} e^{\alpha \hat{Y}} \,.
\eal
${\hat W}(\alpha=1)$ is the left-hand side of the theorem with the initial condition ${\hat W}(\alpha)(\alpha=0) = \hat{I}$.
By taking the derivative with respect to $\alpha$, one finds
\bal
\frac{d}{d\alpha} {\hat W}(\alpha) = \left[ \hat{X} + {\hat W}(\alpha) \hat{Y} {\hat W}^{-1}(\alpha) \right] {\hat W}(\alpha) = \left[ \hat{X} + \hat{Y} + \alpha [\hat{X}, \hat{Y}] \right] {\hat W}(\alpha) \,.
\eal
In the second equality, we have used the BCH lemma and $[\hat{X}, \hat{Y}]$ commutes with $\hat{X}$ and $\hat{Y}$.
One can also introduce
\bal
{\hat W}'(\alpha) = e^{\hat{Z}'(\alpha)} \,, \quad \hat{Z}'(\alpha) = \alpha \hat{X} + \alpha \hat{Y} + \frac{1}{2} \alpha^2 [\hat{X}, \hat{Y}] \,.
\eal
${\hat W}'(\alpha=1)$ is the right-hand side of the theorem with the initial condition ${\hat W}'(\alpha)(\alpha=0) = \hat{I}$.
By taking the derivative with respect to $\alpha$, one finds
\bal
\frac{d}{d\alpha} {\hat W}'(\alpha) = \left[ \hat{X} + \hat{Y} + \alpha [\hat{X}, \hat{Y}] \right] {\hat W}'(\alpha) \,.
\eal
Therefore, both ${\hat W}(\alpha)$ and ${\hat W}'(\alpha)$ satisfy the same ordinary differential equation with the same initial condition. It follows that ${\hat W}(\alpha) = {\hat W}'(\alpha)$ including $\alpha = 1$.

It is known that BCH theorem is generalized even when $[\hat{X}, \hat{Y}]$ does not commutes with $\hat{X}$ and $\hat{Y}$:
\bal
\hat{Z} = \hat{X} + \hat{Y} + \frac{1}{2} [\hat{X}, \hat{Y}] + \frac{1}{12} [\hat{X}, [\hat{X}, \hat{Y}]] + \frac{1}{12} [[\hat{X}, \hat{Y}], \hat{Y}] + \dots \,,
\eal
where dots denote the higher-order commutation relations.
It follows that when $\hat{X}$ and $\hat{Y}$ are elements of a closed algebra, $\hat{Z}$ is also its element.

\section{Derivative of the time-dependent operators} \label{sec:derivative}
Consider the following operator with multiple parameters $\{ \alpha \}$:
\bal
\hat{Y} \left( \{ \alpha \} \right) = \exp \left( \sum_i \alpha_i \hat{X}_i \right) \,.
\eal
Its $\alpha_i$-derivative is given by
\bal
\frac{\partial}{\partial \alpha_i} \hat{Y} \left( \{ \alpha \} \right) = \left[ \int_0^1 dx  \exp \left(x \sum_j \alpha_j \hat{X}_j \right) \hat{X}_i \exp \left(- x \sum_k \alpha_k \hat{X}_k \right) \right] \hat{Y} \left( \{ \alpha \} \right) \,.
\label{eq:derivative}
\eal
Note that the right-hand side cannot be generically reduced to $\hat{X}_i \hat{Y} \left( \{ \alpha \} \right)$, since $\sum_j \alpha_j \hat{X}_j$ does not commute with $\hat{X}_i$.

One can prove it by expanding the exponential in the both sides.
Let us start with the left-hand side:
\bal
\frac{\partial}{\partial \alpha_i} \hat{Y} \left( \{ \alpha \} \right) = \frac{\partial}{\partial \alpha_i} \sum_{n=0}^{\infty} \frac{1}{n!} \left( \sum_j \alpha_j \hat{X}_j \right)^n &= \sum_{n=1}^{\infty} \sum_{m=0}^{n-1} \frac{1}{n!} \left( \sum_j \alpha_j \hat{X}_j \right)^{m} \hat{X}_i \left( \sum_k \alpha_k \hat{X}_k \right)^{n-m-1} \notag \\
&= \sum_{n=0}^{\infty} \sum_{m=0}^{n} \frac{1}{(n+1)!} \left( \sum_j \alpha_j \hat{X}_j \right)^{m} \hat{X}_i \left( \sum_k \alpha_k \hat{X}_k \right)^{n-m} \,.
\eal
Now consider the right-hand side.
By noting
\bal
\int_0^1 dx x^m (1-x)^n = B(m+1, n+1) = \frac{m! n!}{(m+n+1)!} \,,
\eal
one finds
\bal
& \left[ \int_0^1 dx \exp \left(x \sum_j \alpha_j \hat{X}_j \right) \hat{X}_i \exp \left(- x \sum_k \alpha_k \hat{X}_k \right) \right] \hat{Y} \left( \{ \alpha \} \right) \notag \\
&= \int_0^1 dx \exp \left(x \sum_j \alpha_j \hat{X}_j \right) \hat{X}_i \exp \left( (1- x) \sum_k \alpha_k \hat{X}_k \right)  \notag \\
& = \sum_{m,n=0}^{\infty} \frac{1}{m! n!} \int_0^1 dx x^m (1-x)^n \left(\sum_j \alpha_j \hat{X}_j \right)^m \hat{X}_i \left( \sum_k \alpha_k \hat{X}_k \right)^{n} \notag \\
& = \sum_{n,m=0}^{\infty} \frac{1}{(m+n+1)!} \left(\sum_j \alpha_j \hat{X}_j \right)^m \hat{X}_i \left( \sum_k \alpha_k \hat{X}_k \right)^{n} \,.
\eal
Actually, these two expressions are equivalent, since they sum over the two non-negative integers (grids in the $xy$-plane) in different ways: in the left-hand side, $n$ fixes the section on $y$-axis and $m$ goes over each grid along the diagonal line with the slope being $-1$; in the right-hand side, $m$ and $n$ denote $x$ and $y$, respectively. 
By using the BCH lemma, one can also write \eqref{eq:derivative} as
\bal
\frac{\partial}{\partial \alpha_i} \hat{Y} \left( \{ \alpha \} \right) = \left[ \sum_{n=0}^\infty \frac{1}{(n+1)!} \left[\sum_j \alpha_j \hat{X}_j, \hat{X}_i \right]_n \right] \hat{Y} \left( \{ \alpha \} \right) \,.
\eal

Now we are ready to evaluate the time-derivative of the displacement and squeeze operators.
Let us start with the displacement operator:
\bal
\hat{D} \left( \alpha_i(t) \right) = \exp \left[ \alpha_i(t) \hat{a}^\dagger_i - \alpha^*_i(t) \hat{a}_i \right] \,.
\eal
By using \eqref{eq:derivative}, one finds its time derivative as
\bal
&\frac{d}{dt} \hat{D} \left( \alpha_i(t) \right) \notag \\
&= \left[ \int_0^1 dx \exp\left( x \left[ \alpha_i(t) \hat{a}^\dagger_i - \alpha^*_i(t) \hat{a}_i \right] \right) \left( \hat{a}^\dagger_i \frac{d}{dt} \alpha_i(t) - \hat{a}_i \frac{d}{dt} \alpha^*_i(t) \right) \right. \notag \\
& \quad \left. \times \exp\left(- x \left[\alpha_i(t) \hat{a}^\dagger_i - \alpha^*_i(t) \hat{a}_i \right]\right) \right] \hat{D} (\alpha_i(t)) \notag \\
&= \left[ \int_0^1 dx \hat{D} (x \alpha_i(t)) \left( \hat{a}^\dagger_i \frac{d}{dt} \alpha_i(t) - \hat{a}_i \frac{d}{dt} \alpha^*_i(t) \right) \hat{D} (-x \alpha_i(t)) \right] \hat{D} (\alpha_i(t)) \notag \\
& = \left[ \int_0^1 dx \left( \left[ \hat{a}^\dagger_i - x \alpha^*_i(t) \right] \frac{d}{dt} \alpha_i(t) - \left[ \hat{a}_i - x \alpha_i(t) \right] \frac{d}{dt} \alpha_i^*(t) \right) \right] \hat{D} (\alpha_i(t)) \notag \\
& = \left[ \hat{a}^\dagger_i \frac{d}{dt} \alpha_i(t) - \hat{a}_i \frac{d}{dt} \alpha^*_i(t) + \frac{1}{2} \left( \alpha_i(t) \frac{d}{dt} \alpha_i^*(t) - \alpha^*_i(t) \frac{d}{dt} \alpha_i(t) \right) \right] \hat{D} (\alpha_i(t)) \,.
\eal

Then, we consider the squeeze operator:
\bal
\hat{S} \left( \{\zeta(t)\} \right) = \exp\left( \sum_{i,j} \left[ \frac{1}{2} \zeta^*_{ij}(t) \hat{a}_i \hat{a}_j - \frac{1}{2} \zeta_{ij}(t) \hat{a}^\dagger_i \hat{a}^\dagger_j \right] \right)\,.
\eal
By using \eqref{eq:derivative}, one finds its time derivative as
\bal
&\frac{d}{dt} \hat{S} \left( \{\zeta(t)\} \right) \notag \\
&= \left[ \int_0^1 dx \exp\left( x \sum_{k,l} \left[ \frac{1}{2} \zeta^*_{kl}(t) \hat{a}_k \hat{a}_l - \frac{1}{2} \zeta_{kl}(t) \hat{a}^\dagger_k \hat{a}^\dagger_l \right]\right) \sum_{i,j} \left( \frac{1}{2} \hat{a}_i \hat{a}_j \frac{d}{dt} \zeta^*_{ij}(t) - \frac{1}{2} \hat{a}^\dagger_i \hat{a}^\dagger_j \frac{d}{dt} \zeta_{ij}(t) \right) \right. \notag \\
&\quad  \times \exp\left( -x \sum_{m,n} \left[ \frac{1}{2} \zeta^*_{mn}(t) \hat{a}_m \hat{a}_n - \frac{1}{2} \zeta_{mn}(t) \hat{a}^\dagger_m \hat{a}^\dagger_n \right]\right) \hat{S} \left( \{\zeta(t)\} \right) \notag \\
&= \frac{1}{2} \left[  \int_0^1 dx \hat{S} \left( \{x \zeta(t)\} \right) \sum_{i,j} \left( \hat{a}_i \hat{a}_j \frac{d}{dt} \zeta^*_{ij}(t) - \hat{a}^\dagger_i \hat{a}^\dagger_j \frac{d}{dt} \zeta_{ij}(t) \right) \hat{S} \left( \{-x \zeta(t)\} \right) \right] \hat{S} (\{\zeta(t)\}) \notag \\
&= \frac{1}{2} \left( \int_0^1 dx \sum_{i,j,k,l} \left[ \left( \left[ \cosh \left( x [\zeta(t) \zeta^*(t)]^{1/2} \right) \right]_{ik} \hat{a}_k + \left[ \sinh \left( x [\zeta(t) \zeta^*(t)]^{1/2} \right) [\zeta(t) \zeta^*(t)]^{-1/2} \zeta(t) \right]_{ik} \hat{a}^\dagger_k \right) \right. \right. \notag \\
& \quad \times \left( \left[ \cosh \left( x [\zeta(t) \zeta^*(t)]^{1/2} \right) \right]_{jl} \hat{a}_l + \left[ \sinh \left( x [\zeta(t) \zeta^*(t)]^{1/2} \right) [\zeta(t) \zeta^*(t)]^{-1/2} \zeta(t) \right]_{jl} \hat{a}^\dagger_l \right) \frac{d}{dt} \zeta^*_{ij}(t) \notag \\
& \quad - \left( \left[ \cosh \left( x [\zeta^*(t) \zeta(t)]^{1/2} \right) \right]_{ik} \hat{a}^\dagger_k + \left[ \sinh \left(x [\zeta^*(t) \zeta(t)]^{1/2} \right) [\zeta^*(t) \zeta(t)]^{-1/2} \zeta^*(t) \right]_{ik} \hat{a}_k \right) \notag \\
& \left. \left. \quad \times \left( \left[ \cosh \left( x [\zeta^*(t) \zeta(t)]^{1/2} \right) \right]_{jl} \hat{a}^\dagger_l + \left[ \sinh \left( x [\zeta^*(t) \zeta(t)]^{1/2} \right) [\zeta^*(t) \zeta(t)]^{-1/2} \zeta^*(t) \right]_{jl} \hat{a}_l \right) \frac{d}{dt} \zeta_{ij}(t) \right] \right)  \notag \\
& \quad \times \hat{S} \left( \{\zeta(t)\} \right) \,.
\eal
It is not obvious how to perform the $x$-integral in general, and thus in the following we limit our selves within diagonal $\zeta_{ij} = \zeta_{i} \delta_{ij}$.
Then, $\hat{S} \left( \{\zeta(t)\} \right) = \bigotimes_i \hat{S} \left( \zeta_i(t) \right)$
Noting 
\bal
&\int_0^1 dx \sinh^2(x r) = - \frac{1}{2} + \frac{1}{2 r} \sinh(r) \cosh(r) \,, \quad
\int_0^1 dx \cosh^2(x r) = \frac{1}{2} + \frac{1}{2 r} \sinh(r) \cosh(r) \,, \notag \\
&\int_0^1 dx \sinh(x r) \cosh(x r) = \frac{1}{2 r} \sinh^2(r) \,,
\eal
one finds
\bal
&\frac{d}{dt} \hat{S} \left( \zeta_i(t) \right) \notag \\
&= \frac{1}{2} \left[ \int_0^1 dx \left( \left[ \cosh \left( x |\zeta_i(t)| \right) \hat{a}_i + \sinh \left( x |\zeta_i(t)| \right) \frac{\zeta_i(t)}{|\zeta_i(t)|} \hat{a}^\dagger_i \right]^2 \frac{d}{dt} \zeta^*_i(t) \right. \right. \notag \\
& \left. \left. \quad - \left[ \cosh \left( x |\zeta_i(t)| \right) \hat{a}^\dagger_i + \sinh \left(x |\zeta_i(t)| \right) \frac{\zeta^*_i(t)}{|\zeta_i(t)|} \hat{a}_i \right]^2 \frac{d}{dt} \zeta_{i}(t) \right) \right] \hat{S} \left( \zeta_i(t) \right) \notag \\
&= \frac{1}{2} \left[ \int_0^1 dx \left( \cosh^2 \left( x |\zeta_i(t)| \right) \left[ \hat{a}^2_i \frac{d}{dt} \zeta^*_i(t) - \hat{a}^{\dagger 2}_i \frac{d}{dt} \zeta_i(t) \right] + \sinh^2 \left( x |\zeta_i(t)| \right) \left[ \hat{a}^{\dagger 2}_i \frac{\zeta^2_i(t)}{|\zeta_i(t)|^2} \frac{d}{dt} \zeta^*_i(t) \right.  \right. \right. \notag \\
& \left. \left. \left. \quad - \hat{a}^2_i \frac{\zeta^{*2}_i(t)}{|\zeta_i(t)|^2} \frac{d}{dt} \zeta_i(t) \right] + \sinh \left( x |\zeta_i(t)| \right) \cosh \left( x |\zeta_i(t)| \right) \left[ \frac{\zeta_i(t)}{|\zeta_i(t)|} \frac{d}{dt} \zeta^*_i(t) - \frac{\zeta^*_i(t)}{|\zeta_i(t)|} \frac{d}{dt} \zeta_i(t) \right] \left( \hat{a}_i \hat{a}^\dagger_i + \hat{a}^\dagger_i \hat{a}_i \right) \right) \right] \notag \\
& \quad \times \hat{S} \left( \zeta_i(t) \right) \notag \\
&= \frac{1}{2} \left( \left[\frac{1}{2} + \frac{1}{2 |\zeta_i(t)|} \sinh \left( |\zeta_i(t)| \right) \cosh \left( |\zeta_i(t)| \right) \right] \left[ \hat{a}^2_i \frac{d}{dt} \zeta_i^*(t) - \hat{a}^{\dagger 2}_i \frac{d}{dt} \zeta_i(t) \right] \right. \notag \\
&\quad + \left[- \frac{1}{2} + \frac{1}{2 |\zeta_i(t)|} \sinh \left( |\zeta_i(t)| \right) \cosh \left( |\zeta_i(t)| \right) \right] \left[ \hat{a}^{\dagger 2}_i \frac{\zeta^2_i(t)}{|\zeta_i(t)|^2} \frac{d}{dt} \zeta^*_i(t) - \hat{a}^2_i \frac{\zeta^{*2}_i(t)}{|\zeta_i(t)|^2} \frac{d}{dt} \zeta_i(t) \right] \notag \\
&\quad \left. + \frac{1}{2 |\zeta_i(t)|} \sinh^2 \left( |\zeta_i(t)| \right) \left[ \frac{\zeta_i(t)}{|\zeta_i(t)|} \frac{d}{dt} \zeta^*_i(t) - \frac{\zeta^*_i(t)}{|\zeta_i(t)|} \frac{d}{dt} \zeta_i(t) \right] \left( \hat{a}_i \hat{a}^\dagger_i + \hat{a}^\dagger_i \hat{a}_i \right) \right) \hat{S} \left( \zeta_i(t) \right) \notag\\
&=\frac{1}{4} \left( \left[ \zeta^*_i(t) \hat{a}^2_i - \zeta_i(t) \hat{a}^{\dagger 2}_i \right] \left[ \frac{\zeta_i(t)}{|\zeta_i(t)|^2} \frac{d}{dt} \zeta^*_i(t) + \frac{\zeta^*_i(t)}{|\zeta_i(t)|^2} \frac{d}{dt} \zeta_i(t) \right] + \left[ \frac{\zeta_i(t)}{|\zeta_i(t)|} \sinh \left( |\zeta_i(t)| \right) \cosh \left( |\zeta_i(t)| \right) \hat{a}^{\dagger 2}_i  \right. \right. \notag \\
&\left. \left. \quad + \frac{\zeta^*_i(t)}{|\zeta_i(t)|} \sinh \left( |\zeta_i(t)| \right) \cosh \left( |\zeta_i(t)| \right) \hat{a}^2_i + \sinh^2 \left( |\zeta_i(t)| \right) \left( \hat{a}_i \hat{a}^\dagger_i + \hat{a}^\dagger_i \hat{a}_i \right)  \right]  \left[ \frac{\zeta_i(t)}{|\zeta_i(t)|^2} \frac{d}{dt} \zeta^*_i(t) - \frac{\zeta^*_i(t)}{|\zeta_i(t)|^2} \frac{d}{dt} \zeta_i(t) \right] \right)  \notag \\
& \quad \times \hat{S} \left( \zeta_i(t) \right) \notag \\
&= \frac{1}{4} \left(  \left[ \zeta^*_i(t) \hat{a}^2_i - \zeta_i(t) \hat{a}^{\dagger 2}_i \right] \frac{d}{dt} \ln \left( |\zeta_i(t)|^2 \right) + \left[ \hat{S} \left( \{\zeta(t)\} \right) \hat{a}^\dagger_i \hat{a}_i \hat{S} \left( \{-\zeta(t)\} \right) - \hat{a}^\dagger_i \hat{a}_i \right] \frac{d}{dt} \ln \left( \zeta^*_i(t) / \zeta_i(t) \right) \right)  \notag \\
& \quad \times \hat{S} \left( \zeta_i(t) \right) \,.
\eal


\section{Check of numerical convergence}\label{sec:num_conv}
\begin{figure}[H]
    \centering
    \includegraphics[width=.6\linewidth]{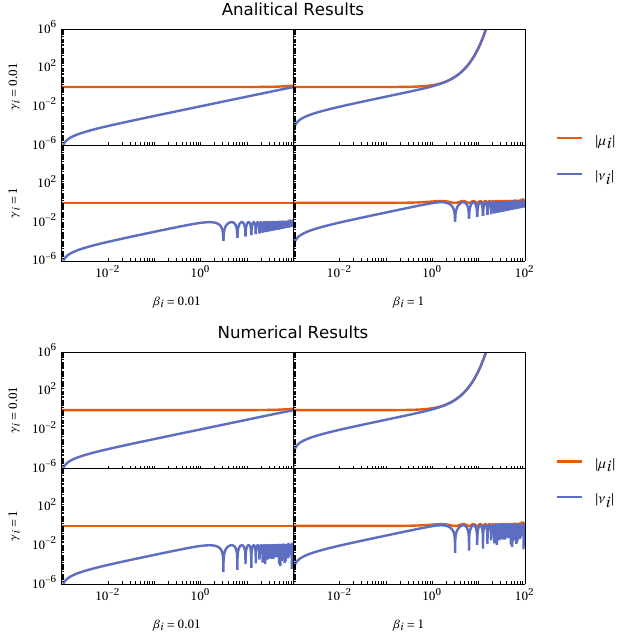}
    \caption{Comparison between the analytical (top) and numerical (bottom) time evolution of $|\mu_i (t)|$ and $|\nu_i (t)|$. Here we take values of $\beta_i, \gamma_i = 0.01, 1$.}
    \label{fig:sanitycheck}
\end{figure}

The following section demonstrates the convergence of our numerical method solving the differential equation \eqref{eq:matrix-form}.
We consider the case of $\{h(t)\} = \{f(t)\} = 0$ and ``real'' diagonal $g_{ij}(t) = g_i(t) \delta_{ij}$ where the exact solution is available as discussed in section \ref{sec:realg}.
To be concrete, we consider
\bal
g_i(t)=\beta_i \cos (\gamma_i t) \,,
\eal
where $\beta_i$ and $\gamma_i$ are real constants.
Correspondingly,
\bal
\mu_i(t)=\cosh \left(\frac{\beta_i \sin (\gamma_i  t)}{\gamma_i }\right) \,, \quad \nu_i(t)=\sinh \left(\frac{\beta_i \sin (\gamma_i  t)}{\gamma_i }\right) \,.
\eal
The comparison is shown in figure \ref{fig:sanitycheck}.
As can be seen in the figure, the numerical solution converges to the exact one at any time $t$, which confirms the convergence of our numerical methods.

\acknowledgments
A. K. thanks Takumi Kuwahara and Kodai Sakurai for valuable comments on the manuscript.
A. K. acknowledges partial support from Norwegian Financial Mechanism for years 2014-2021, grant nr 2019/34/H/ST2/00707; and from National Science Centre, Poland, grant DEC-2018/31/B/ST2/02283.
A. M. acknowledges the use of ChatGPT (OpenAI) for assistance with English grammar and language editing.
All scientific content, analysis, and conclusions are the sole responsibility of the authors.

\bibliography{biblio} 
\bibliographystyle{JHEP}

\end{document}